\documentclass[10pt,a4paper]{article}

\usepackage{cite}
\usepackage[latin1]{inputenc}
\usepackage{graphicx}
\usepackage{amsmath}
\usepackage{amsfonts}
\usepackage{amssymb}
\usepackage{makeidx}
\usepackage{setspace}
\usepackage{geometry}
\usepackage{subfigure}
\usepackage{color}
\usepackage{epstopdf}

\usepackage{lineno}

\title{Neural Population Coding of Multiple Stimuli}
\author{A. Emin Orhan$^{\dagger}$ \hspace{5mm}  Wei Ji Ma$^{\dagger,\ddagger}$\\
$\dagger$Center for Neural Science and $\ddagger$Department of Psychology \\
New York University, New York, NY 10003\\
\\
\\
\\
{Abbreviated title:} Neural Population Coding of Multiple Stimuli \\
{Corresponding author:} A. Emin Orhan \\
{E-mail:} \texttt{eorhan@cns.nyu.edu} \\
{Address:} Center for Neural Science, 4 Washington Place, New York NY 10003 \\
\\
\\
\\
Number of pages: 43\\
Number of figures: 12\\
Number of tables: 1 \\
Number of words (Abstract): 218\\
Number of words (Introduction): 496 \\
Number of words (Discussion): 1410
}

\begin{document}

\maketitle

\vspace{\fill} \footnoterule \noindent \textbf{Acknowledgments:} WJM was supported by grant R01EY020958 from the National Eye Institute and grant W911NF-12-1-0262 from the Army Research Office. The authors declare no competing financial interests.

\clearpage
\begin{abstract}
{In natural scenes, objects generally appear together with other objects. Yet, theoretical studies of neural population coding typically focus on the encoding of single objects in isolation. Experimental studies suggest that neural responses to multiple objects are well described by linear or nonlinear combinations of the responses to constituent objects, a phenomenon we call stimulus mixing. Here, we present a theoretical analysis of the consequences of common forms of stimulus mixing observed in cortical responses. We show that some of these mixing rules can severely compromise the brain's ability to decode the individual objects. This cost is usually greater than the cost incurred by even large reductions in the gain or large increases in neural variability, explaining why the benefits of attention can be understood primarily in terms of a stimulus selection, or demixing, mechanism rather than purely as a gain increase or noise reduction mechanism. The cost of stimulus mixing becomes even higher when the number of encoded objects increases, suggesting a novel mechanism that might contribute to set size effects observed in myriad psychophysical tasks. We further show that a specific form of neural correlation and heterogeneity in stimulus mixing among the neurons can partially alleviate the harmful effects of stimulus mixing. Finally, we derive simple conditions that must be satisfied for unharmful mixing of stimuli.}
\end{abstract}

\clearpage
\section*{Introduction}
In natural vision, objects typically appear within the context of other objects rather than in isolation. It is, therefore, important to understand how cortical neurons encode multiple objects. Experimental studies suggest that in many cortical areas, neural responses to the presentation of multiple stimuli can be successfully described as a linear or nonlinear combination of the responses to the individual stimuli. In area IT, for example, responses of many individual neurons to pairs and triplets of objects are well described by the average of their responses to individual stimuli (Zoccolan et al., 2005; Zoccolan et al., 2007). A similar weighted averaging model also provides a good description of the responses of motion-selective MT (Recanzone et al., 1997; Britten and Heuer, 1999) and MST neurons (Recanzone et al., 1997) to pairs of moving objects, responses of V4 neurons to composite shapes consisting of several oriented line segments (Nandy et al., 2013), population responses in V1 to simultaneously presented gratings (Busse et al., 2009; MacEvoy et al., 2009), and at a larger scale, fMRI responses to multiple objects in object selective area LOC (MacEvoy and Epstein, 2009) and in V4 (Beck and Kastner, 2007). 

In working memory and associative learning tasks, when multiple stimuli have to be stored in memory simultaneously, responses of single neurons in prefrontal cortex are again a potentially complex function of multiple stimuli as well as other task parameters (Duncan, 2001; Warden and Miller, 2007; Warden and Miller, 2010). Such ``mixed selectivity'' has been argued to be crucial for successful performance in context-dependent behavioral tasks (Rigotti et al., 2010; Rigotti et al., 2013). However, mixed selectivity is not unreservedly beneficial (Barak et al., 2013). By mapping two similar points in the input space to points that are farther apart in the output space, stimulus mixing can make them more easily discriminable. The same, however, applies to noisy versions of the same stimulus that one would not want to make more discriminable, thus creating a problem of generalization or robustness against noise (Barak et al., 2013). The extent of this problem for commonly observed forms of stimulus mixing in the brain is unknown and an analysis of what types of mixing are more or less vulnerable to this problem is lacking.                  

In this article, using both analytical and numerical tools, we present a systematic analysis of some common forms of stimulus mixing observed in cortical responses with regard to their consequences for stimulus encoding in the presence of neural variability. We show that some of these common mixing rules, such as weighted averaging, can be profoundly harmful for stimulus encoding. Another commonly observed, divisive form of stimulus mixing (Allman et al., 1985; Cavanaugh et al., 2002) can also be harmful for stimulus encoding, although much less so than weighted averaging. We also derive mathematical conditions that must be satisfied for unharmful mixing of stimuli, and provide geometric explanations for what makes particular forms of stimulus mixing more or less harmful than others. 

\section*{Materials and Methods}
\subsection*{Derivation of the Fisher information matrix} \label{derivation_sec}
\begin{table}
\centering 
\begin{tabular}{ll}
\hline
\textbf{Symbol}    	     & \textbf{Meaning} \\
\hline
$n$          			 & Number of neurons per group    						   						\\
$I_{ij}$     			 & $ij$-th term of the Fisher information matrix           						\\
$\mathbf{f}$ 		     & Column vector of the mean responses of all neurons      						\\
$s_i$        		     & $i$-th stimulus     									   						\\
$\Delta s = |s_1 - s_2|$ & Angular distance between two stimuli      									\\
$\phi_k$     			 & Preferred stimulus of $k$-th neuron      			   						\\
$R$          			 & Correlation matrix									   						\\
$S$    		 		     & Diagonal matrix of standard deviations 				   						\\
$Q$, $\Sigma$ 			 & Covariance matrix of neural responses										\\
$\alpha$     			 & Gain of tuning functions (Equation~\ref{vonmises})      						\\
$c_0$        		     & Maximum noise correlation between neurons (Equation~\ref{R_matrix}) 			\\
$L$           			 & Correlation length scale parameter (Equation~\ref{R_matrix}) 				\\
$\beta$      			 & Scaling factor for across-group correlations (Equation~\ref{R_prime_matrix}) \\
$w$ 		 			 & Mixing weight in the linear mixing model (Equation~\ref{linear_mixing_mean}) \\
$\sigma_w^2$ 			 & Variance of the distribution over mixing weights 							\\
$U$						 & The unitary discrete Fourier transform (DFT) matrix 							\\
$\tilde{\mathbf{a}}$     & DFT of a vector $\mathbf{a}$													\\
$\breve{\mathbf{a}}$     & Inverse DFT of a vector $\mathbf{a}$											\\
$N$  		 		     & Set size (number of encoded stimuli) 										\\ 
$\nu$  		 		     & Exponent in the nonlinear mixing model of Britten and Heuer (Equation~\ref{brittenheuer99_eq}) \\
$k_w$ 					 & Divisive normalization scaling factor (Equation~\ref{div_norm_mean_resp_eq})  \\     					
$\mathbf{J}$             & Jacobian matrix for the mean responses of the neurons (Equation~\ref{jacobian_responses}) \\
$K = \mathrm{Tr}[\mathbf{J}^T \mathbf{J}]$ & Total resource, i.e. the sum of the squares of the derivatives in $\mathbf{J}$ \\																	
\hline
\end{tabular} 
\caption{List of frequently used symbols.}
\end{table}
We use a multivariate Gaussian distribution to model neural variability. For a Gaussian distribution, the $ij$-th term of the Fisher information matrix (FIM) is given by (Abbott and Dayan, 1999): 
\begin{eqnarray}
I_{ij} & = & I_{ij,\mathrm{mean}} + I_{ij,\mathrm{cov}} \nonumber \\
	   & = & \frac{\partial \mathbf{f}^T}{\partial s_i} Q^{-1} \frac{\partial \mathbf{f}}{\partial s_j} + \frac{1}{2} \mathrm{Tr}\bigg[Q^{-1}\frac{\partial Q}{\partial s_i}Q^{-1}\frac{\partial Q}{\partial s_j}\bigg],
\end{eqnarray} 
where $\mathbf{f}$ is a column vector of the mean responses of all neurons in the population. In the linear mixing model (see {\it Results}), the mean response of a neuron $k$ to a pair of stimuli $(s_1,s_2)$ is assumed to be a weighted average of its mean responses to each individual stimulus alone (Equation~\ref{linear_mixing_mean}). The individual tuning functions describing the mean responses of neurons to single stimuli are assumed to be von Mises:
\begin{equation}
f(s;\phi) = \alpha \exp\big(\gamma  [\cos(s-\phi) - 1]\big) + \eta,
\label{vonmises}
\end{equation}
where $\alpha$, $\gamma$ and $\eta$ are the tuning function parameters and $\phi$ is the neuron's prefered stimulus. Here and in the rest of the paper, differences between circular variables should always be understood as angular differences. The covariance matrix $Q$ can be expressed as $Q = S R S$, where $S$ is a diagonal matrix of the standard deviations of neural responses and $R$ is the correlation matrix. In our problem, $R$ has a block structure:
\begin{equation}
R = \begin{bmatrix}
A & B \\
B & A 
\end{bmatrix}
\label{R_block_matrix}
\end{equation}
with $A$ representing the correlations between the neurons within the same group and $B$ representing the across-group correlations. We assume that within-group correlations decay exponentially with the angular difference between the prefered stimuli of neurons:
\begin{equation}
A_{km} = \delta_{km} + (1-\delta_{km})c_0\exp\bigg(-\frac{|\phi_k - \phi_m|}{L}\bigg),
\label{A_matrix}
\end{equation}
where $\delta$ is the Kronecker delta function. Across-group correlations are simply scaled versions of the within-group correlations:
\begin{equation}
B_{km^\prime} = \beta c_0 \exp\bigg(-\frac{|\phi_k - \phi_{m^\prime}|}{L}\bigg).
\label{B_matrix}
\end{equation}

The inverse of the covariance matrix is given by $Q^{-1} = S^{-1} R^{-1} S^{-1}$. Since $S$ is diagonal, its inverse is straightforward. The inverse of $R$ is less so. From Equation~\ref{R_block_matrix}, blockwise inversion of $R$ yields:
\begin{equation}
R^{-1} = \begin{bmatrix}
(A-BA^{-1}B)^{-1} & -A^{-1}B(A-BA^{-1}B)^{-1} \\
 -A^{-1}B(A-BA^{-1}B)^{-1} & (A-BA^{-1}B)^{-1}
\end{bmatrix}.
\label{R_inverse}
\end{equation}
Importantly, $A$ and $B$ are circulant matrices, hence they are both diagonalized in the Fourier basis. This implies that Equation~\ref{R_inverse} can be written as:
\begin{equation}
R^{-1} = \begin{bmatrix}
U & 0 \\
0 & U
\end{bmatrix}
\begin{bmatrix}
(\tilde{A}-\tilde{B}\tilde{A}^{-1}\tilde{B})^{-1} & -\tilde{A}^{-1}\tilde{B}(\tilde{A}-\tilde{B}\tilde{A}^{-1}\tilde{B})^{-1} \\
 -\tilde{A}^{-1}\tilde{B}(\tilde{A}-\tilde{B}\tilde{A}^{-1}\tilde{B})^{-1} & (\tilde{A}-\tilde{B}\tilde{A}^{-1}\tilde{B})^{-1}
\end{bmatrix}
\begin{bmatrix}
U^{\ast} & 0 \\
0 & U^{\ast}
\end{bmatrix},
\end{equation}
where $U$ is the unitary discrete Fourier transform matrix with entries $U_{kj} = \exp(-2\pi i k j /n) / \sqrt{n}$ (where $n$ is the number of neurons in each group), $U^{\ast}$ is its conjugate transpose, and $\tilde{A}$ and $\tilde{B}$ are diagonal matrices of eigenvalues of $A$ and $B$ respectively which can be found by taking the discrete Fourier transforms (DFT) of the first columns of $A$ and $B$. 

Let us denote $\tilde{\mathbf{a}} = \mathrm{diag}(\tilde{A})$ and $\tilde{\mathbf{b}} = \mathrm{diag}(\tilde{B})$ to be the diagonals of $\tilde{A}$ and $\tilde{B}$. Note that because $\tilde{A}$ and $\tilde{B}$ are diagonal matrices:
\begin{equation}
\bigg((\tilde{A}-\tilde{B}\tilde{A}^{-1}\tilde{B})^{-1}\bigg)_{kk} = C_{kk} = \frac{1}{\tilde{\mathbf{a}}_k - \frac{\tilde{\mathbf{b}}_k^2}{\tilde{\mathbf{a}}_k}} = \frac{\tilde{\mathbf{a}}_k}{\tilde{\mathbf{a}}_k^2 - \tilde{\mathbf{b}}_k^2}.
\label{C_nn}
\end{equation}    
Similarly:
\begin{eqnarray}
\bigg(-\tilde{A}^{-1}\tilde{B}(\tilde{A}-\tilde{B}\tilde{A}^{-1}\tilde{B})^{-1}\bigg)_{kk} = D_{kk} = -\frac{\tilde{\mathbf{b}}_k}{\tilde{\mathbf{a}}_k^2 - \tilde{\mathbf{b}}_k^2}. 
\label{D_nn}
\end{eqnarray}

\subsection*{Poisson-like noise} 
We first derive $I_{\mathrm{mean}}$ and $I_{\mathrm{cov}}$ for a Poisson-like noise model where the mean responses of neurons are equal to their variance.

{\bf Derivation of $I_{\mathrm{mean}}$:} We can write down the first term of the Fisher information matrix as follows:
\begin{equation}
I_{ij,\mathrm{mean}} = \frac{\partial \mathbf{f}^T}{\partial s_i} Q^{-1} \frac{\partial \mathbf{f}}{\partial s_j} = \frac{\partial \mathbf{f}^T}{\partial s_i} \begin{bmatrix}
S_1 & 0 \\
0 & S_2
\end{bmatrix}^{-1}
\begin{bmatrix}
U & 0 \\
0 & U
\end{bmatrix}
\begin{bmatrix}
C & D \\
D & C
\end{bmatrix}
\begin{bmatrix}
U^{\ast} & 0 \\
0 & U^{\ast}
\end{bmatrix}
\begin{bmatrix}
S_1 & 0 \\
0 & S_2
\end{bmatrix}^{-1}
\frac{\partial \mathbf{f}}{\partial s_j},
\label{Imean_long}
\end{equation}
where $S_1$ and $S_2$ are diagonal matrices of the standard deviations $\sigma_k$ of the responses of neurons in the first and second group respectively. For Poisson-like noise, $\sigma_k = \sqrt{f_k}$. In the following we denote by $\mathbf{g}^i$ the vector whose $k$-th entry is equal to $\sigma_k^{-1}\partial f_k / \partial s_i$, i.e. the derivative of the $k$-th neuron's mean response with respect to $s_i$ divided by the standard deviation of its variability, where $k$ ranges only over the neurons in the first group. Similarly, we denote by $\mathbf{h}^i$ the vector whose $k$-the entry is equal to $\sigma_k^{-1}\partial f_k / \partial s_i$, but where $k$ now ranges over the neurons in the second group only.

With this notation, we can re-write Equation~\ref{Imean_long} as follows:
\begin{equation}
I_{ij,\mathrm{mean}}  =  (\tilde{\mathbf{g}}^i)^T C \breve{\mathbf{g}}^j + (\tilde{\mathbf{h}}^i)^T D \breve{\mathbf{g}}^j + (\tilde{\mathbf{g}}^i)^T D \breve{\mathbf{h}}^j + (\tilde{\mathbf{h}}^i)^T C \breve{\mathbf{h}}^j ,
\label{dft_version}
\end{equation}
where $\tilde{\mathbf{g}}^i = \sqrt{n} U \mathbf{g}^i$ and $\breve{\mathbf{g}}^i = U^\ast \mathbf{g}^i / \sqrt{n}$ represent the DFT and the inverse DFT of $\mathbf{g}^i$ respectively. Similarly, $\tilde{\mathbf{h}}^i$ and $\breve{\mathbf{h}}^i$ are the DFT and the inverse DFT of $\mathbf{h}^i$. Recall also that $C$ and $D$ are diagonal matrices defined in Equations~\ref{C_nn} and \ref{D_nn} respectively. Note that there are different conventions on how to compute the DFT and the inverse DFT; our usage is consistent with Matlab's implementation of \texttt{fft} and \texttt{ifft} functions. 

The scaling of $I_{ij,\mathrm{mean}}$ with $n$ is similar to the corresponding scaling relationship in the case of the encoding of a single stimulus analyzed previously in Sompolinsky et al. (2001) and in Ecker et al. (2011): for a homogeneous population, Equation~\ref{dft_version} saturates to a finite value in the presence of noise correlations ($c_0 \neq 0$), but diverges for an independent population ($c_0 = 0$). A detailed analysis of the asymptotic behavior of $I_{\mathrm{mean}}$ is provided below for the simpler case of additive noise. 

{\bf Derivation of $I_{\mathrm{cov}}$:} We now derive the second term of the FIM, $I_{ij,\mathrm{cov}}$. We first recall that $Q = SRS$ and then note that $\frac{\partial Q}{\partial s_i} = \partial_i S RS + SR \partial_i S$ where we use the shorthand notation $\partial_i S$ to denote $\frac{\partial S}{\partial s_i}$. We then have: \small
\begin{eqnarray} 
I_{ij,\mathrm{cov}} & = & \frac{1}{2} \mathrm{Tr}\bigg[Q^{-1}\frac{\partial Q}{\partial s_i}Q^{-1}\frac{\partial Q}{\partial s_j}\bigg] = \frac{1}{2} \mathrm{Tr}\bigg[ \big(S^{-1}R^{-1}S^{-1}\partial_i S R S + S^{-1}\partial_i S\big) \big(S^{-1}R^{-1}S^{-1} \partial_j S R S + S^{-1} \partial_j S \big) \bigg] \nonumber \\
& = & \frac{1}{2} \bigg( \mathrm{Tr}\big[ S^{-1} \partial_i S S^{-1} \partial_j S \big] + \mathrm{Tr}\big[ S^{-1}RS^{-1} \partial_i S R \partial_j S \big] + \mathrm{Tr}\big[S^{-1}RS^{-1} \partial_j S R \partial_i S \big] + \mathrm{Tr}\big[ S^{-1} \partial_i S S^{-1} \partial_j S \big] \bigg),
\label{I_cov_expanded}
\end{eqnarray} \normalsize
where in the second line we used the fact that the trace operator is invariant under cyclic permutations of the products of matrices. We now note that $S^{-1} \partial_i S S^{-1} \partial_j S$ is a diagonal matrix and its trace is given by:
\begin{equation}
\mathrm{Tr}\big[ S^{-1} \partial_i S S^{-1} \partial_j S \big] = \sum_{k=0}^{n-1} \mathbf{p}_k^i \mathbf{p}_k^j + \sum_{k=0}^{n-1} \mathbf{t}_k^i \mathbf{t}_k^j , 
\end{equation}
where we introduced the notation $\mathbf{p}^i$ for the vector consisting of the diagonal entries of $S_1^{-1} \partial_i S_1$ and $\mathbf{t}^i$ for the diagonal of $S_2^{-1} \partial_i S_2$. For the second term on the right hand side in Equation~\ref{I_cov_expanded}, we have:
\begin{eqnarray}
\mathrm{Tr}\big[ S^{-1}RS^{-1} \partial_i S R \partial_j S \big] & = & \mathrm{Tr} \big[ U^\ast V_1^i U \tilde{A} U^\ast V_1^j U C \big] + \mathrm{Tr} \big[ U^\ast V_1^i U \tilde{B} U^\ast V_2^j U D \big] + \mathrm{Tr} \big[ U^\ast V_2^i U \tilde{B} U^\ast V_1^j U D \big] \nonumber \\
& + & \mathrm{Tr} \big[ U^\ast V_2^i U \tilde{A} U^\ast V_2^j U C \big]  .
\label{four_trace_terms}
\end{eqnarray}
where we denote $V_1^i = S_1^{-1} \partial_i S_1$ and $V_2^i = S_2^{-1} \partial_i S_2$. Note that the diagonals of $V_1^i$ and $V_2^i$ are the vectors $\mathbf{p}^i$ and $\mathbf{t}^i$. 

Considering the first term on the right hand side in Equation~\ref{four_trace_terms}, we can write it as follows:
\begin{eqnarray}
\mathrm{Tr} \big[ U^\ast V_1^i U \tilde{A} U^\ast V_1^j U C \big] & = &  \sum_{k,l,m,o} U_{kl}^\ast [V_1^i]_{ll} U_{lm} \tilde{A}_{mm} U_{mo}^\ast [V_1^j]_{oo} U_{ok} C_{kk} \nonumber \\
& = & \frac{1}{n^2} \sum_{m,k} \tilde{\mathbf{p}}_{m-k}^i \tilde{A}_{mm} (\tilde{\mathbf{p}}_{m-k}^j)^\ast C_{kk} \nonumber \\
& = & \frac{1}{n^2} \sum_{k=0}^{n-1} C_{kk} \big[ \mathrm{conv}(\tilde{\mathbf{p}}^i\circ (\tilde{\mathbf{p}}^j)^\ast, \tilde{\mathbf{a}}) \big]_k ,
\end{eqnarray}
where $\circ$ is the Hadamard (element-wise) product and $\mathrm{conv}(\cdot,\cdot)$ denotes the circular convolution of two vectors. From the first to the second line above, we simply used the definition of the DFT of a vector. The third line follows from the definition of the circular convolution. For vectors, we use $(\cdot)^\ast$ to denote element-wise conjugation (without transposition). We can express the other terms in Equation~\ref{four_trace_terms} in a similar fashion. Thus, Equation~\ref{four_trace_terms} becomes:
\begin{eqnarray}
\mathrm{Tr}\big[ S^{-1}RS^{-1} \partial_i S R \partial_j S \big] & = & \frac{1}{n^2} \bigg[ \sum_{k=0}^{n-1} C_{kk} \big[ \mathrm{conv}(\tilde{\mathbf{p}}^i\circ (\tilde{\mathbf{p}}^j)^\ast, \tilde{\mathbf{a}}) \big]_k + \sum_{k=0}^{n-1} C_{kk} \big[ \mathrm{conv}(\tilde{\mathbf{p}}^i\circ (\tilde{\mathbf{t}}^j)^\ast, \tilde{\mathbf{b}}) \big]_k \nonumber \\
& + & \sum_{k=0}^{n-1} D_{kk} \big[ \mathrm{conv}(\tilde{\mathbf{t}}^i\circ (\tilde{\mathbf{p}}^j)^\ast, \tilde{\mathbf{b}}) \big]_k + \sum_{k=0}^{n-1} C_{kk} \big[ \mathrm{conv}(\tilde{\mathbf{t}}^i\circ (\tilde{\mathbf{t}}^j)^\ast, \tilde{\mathbf{a}}) \big]_k \bigg] . \label{srssrs_eq}
\end{eqnarray}
We can derive a similar expression for the third term in Equation~\ref{I_cov_expanded}:
\begin{eqnarray}
\mathrm{Tr}\big[ S^{-1}RS^{-1} \partial_j S R \partial_i S \big] & = & \frac{1}{n^2} \bigg[ \sum_{k=0}^{n-1} C_{kk} \big[ \mathrm{conv}(\tilde{\mathbf{p}}^j\circ (\tilde{\mathbf{p}}^i)^\ast, \tilde{\mathbf{a}}) \big]_k + \sum_{k=0}^{n-1} C_{kk} \big[ \mathrm{conv}(\tilde{\mathbf{p}}^j\circ (\tilde{\mathbf{t}}^i)^\ast, \tilde{\mathbf{b}}) \big]_k \nonumber \\
& + & \sum_{k=0}^{n-1} D_{kk} \big[ \mathrm{conv}(\tilde{\mathbf{t}}^j\circ (\tilde{\mathbf{p}}^i)^\ast, \tilde{\mathbf{b}}) \big]_k + \sum_{k=0}^{n-1} C_{kk} \big[ \mathrm{conv}(\tilde{\mathbf{t}}^j\circ (\tilde{\mathbf{t}}^i)^\ast, \tilde{\mathbf{a}}) \big]_k \bigg] . \label{srssrs_eq_2}
\end{eqnarray}

We now note that the first term on the right hand side of Equation~\ref{I_cov_expanded} is equal to the last term. Thus, putting it all together, $I_{ij,\mathrm{cov}}$ can be written as:
\begin{eqnarray}
I_{ij,\mathrm{cov}} & = & \sum_{k=0}^{n-1} \mathbf{p}_k^i \mathbf{p}_k^j + \sum_{k=0}^{n-1} \mathbf{t}_k^i \mathbf{t}_k^j + \frac{1}{2n^2} \bigg[ \sum_{k=0}^{n-1} C_{kk} \big[ \mathrm{conv}(\tilde{\mathbf{p}}^i\circ (\tilde{\mathbf{p}}^j)^\ast, \tilde{\mathbf{a}}) \big]_k + \sum_{k=0}^{n-1} C_{kk} \big[ \mathrm{conv}(\tilde{\mathbf{p}}^i\circ (\tilde{\mathbf{t}}^j)^\ast, \tilde{\mathbf{b}}) \big]_k \nonumber \\
& + & \sum_{k=0}^{n-1} D_{kk} \big[ \mathrm{conv}(\tilde{\mathbf{t}}^i\circ (\tilde{\mathbf{p}}^j)^\ast, \tilde{\mathbf{b}}) \big]_k + \sum_{k=0}^{n-1} C_{kk} \big[ \mathrm{conv}(\tilde{\mathbf{t}}^i\circ (\tilde{\mathbf{t}}^j)^\ast, \tilde{\mathbf{a}}) \big]_k \bigg] \nonumber \\
& + & \frac{1}{2n^2} \bigg[ \sum_{k=0}^{n-1} C_{kk} \big[ \mathrm{conv}(\tilde{\mathbf{p}}^j\circ (\tilde{\mathbf{p}}^i)^\ast, \tilde{\mathbf{a}}) \big]_k + \sum_{k=0}^{n-1} C_{kk} \big[ \mathrm{conv}(\tilde{\mathbf{p}}^j\circ (\tilde{\mathbf{t}}^i)^\ast, \tilde{\mathbf{b}}) \big]_k \nonumber \\
& + & \sum_{k=0}^{n-1} D_{kk} \big[ \mathrm{conv}(\tilde{\mathbf{t}}^j\circ (\tilde{\mathbf{p}}^i)^\ast, \tilde{\mathbf{b}}) \big]_k + \sum_{k=0}^{n-1} C_{kk} \big[ \mathrm{conv}(\tilde{\mathbf{t}}^j\circ (\tilde{\mathbf{t}}^i)^\ast, \tilde{\mathbf{a}}) \big]_k \bigg] . \label{I_ij_cov_dft}
\end{eqnarray}

As for $I_{ij,\mathrm{mean}}$, the scaling of $I_{ij,\mathrm{cov}}$ with $n$ is identical to the corresponding scaling relationship studied in Ecker et al. (2011) for the case of the encoding of a single stimulus: asymptotically $I_{ij,\mathrm{cov}}$ scales linearly with $n$ regardless of the amount of correlations in the population.  

{\bf Effects of heterogeneity in mixing weights in the linear mixing model on $I_{\mathrm{mean}}$ and $I_{\mathrm{cov}}$:} For Poisson-like noise, it is difficult to analytically quantify the effect of heterogeneity in mixing weights on $I_{\mathrm{mean}}$. Considering a single neuron $k$, when there is heterogeneity in mixing weights, we have:
\begin{equation}
\mathbf{g}_k^i = \sigma_k^{-1} \partial f_k / \partial s_i = \frac{(w_i + \delta w_i) f^\prime(s_i; \phi_k) }{\sqrt{(w_i + \delta w_i) f(s_i; \phi_k) + (w_{-i} + \delta w_{-i}) f(s_{-i}; \phi_k)}} ,
\end{equation}  
where we use $\delta w_i$ and $\delta w_{-i}$ to denote the random fluctuations around the mean mixing weights (the subscript $-i$ indicates the stimulus that is not the $i$-th stimulus). In this expression, it is not possible to separate out the effect of mixing weights, as it is in the case of the encoding of a single stimulus studied in Ecker et al. (2011). This makes it difficult to compute expectations over the random fluctuations of mixing weights.

Similarly, unlike in Ecker et al. (2011), heterogeneity in mixing weights also affects the $I_{\mathrm{cov}}$ term in our model. Again, this is because the $k$-th diagonal entry of $S_1^{-1}\partial_i S_1$ is of the following form (a similar expression holds for the diagonal entries of $S_2^{-1} \partial_i S_2$):
\begin{equation}
\mathbf{p}_k^i = \frac{1}{2}\frac{(w_i+ \delta w_i)f^\prime(s_i; \phi_k)}{(w_i+ \delta w_i)f(s_i; \phi_k) + (w_{-i}+ \delta w_{-i})f(s_{-i}; \phi_k)}. \label{s_inv_s_partial}
\end{equation}
The weights in the numerator and the denominator do not cancel in this expression, as they do in the case of the encoding of a single stimulus (Ecker et al., 2011). Unfortunately, because the random fluctuations appear in divisive form in the above expressions (and inside a further square root nonlinearity in the expression for $\mathbf{g}_k^i$), it is difficult to quantify the effect of heterogeneity in mixing weights on $I_{\mathrm{mean}}$ and $I_{\mathrm{cov}}$. Moreover, the asymptotic variance of the optimal estimator, in its turn, is related to the terms of the FIM through an additional nonlinearity (Equation~\ref{var_analytic} below).

Because of these difficulties, we resort to sampling to quantitatively account for the effect of heterogeneity in mixing weights on $I_{\mathrm{mean}}$ and $I_{\mathrm{cov}}$ and on the asymptotic variance of the optimal estimator. In particular, we draw the mixing weights of the neurons independently from beta distributions with mean $w_1$ or $w_2$ and variance $\sigma_w^2$. The weights are then plugged in Equation~\ref{dft_version} for $I_{\mathrm{mean}}$ and in Equation~\ref{I_ij_cov_dft} for $I_{\mathrm{cov}}$. {\color{black}A mathematical analysis of the effect of heterogeneity in mixing weights on the FIM is given below for the simpler case of additive noise.}  

{\bf Asymptotic variance and correlation of optimal unbiased estimates:} After computing $I_{\mathrm{mean}}$ and $I_{\mathrm{cov}}$, we find the asymptotic variance of the optimal unbiased estimates of the individual stimuli and the correlation between the two estimates as follows:
\begin{eqnarray}
\mathrm{Var}(\hat{s}_1|\mathbf{s}) & = & \frac{I_{22}}{I_{11}I_{22} - I_{12}I_{21}}  = \frac{I_{11}}{I_{11}^2 - I_{12}^2} \label{var_analytic}\\
\mathrm{Corr}(\hat{s}_1,\hat{s}_2|\mathbf{s}) & = & -\frac{I_{12}}{\sqrt{I_{11}I_{22}}} = -\frac{I_{12}}{I_{11}} , \label{corr_analytic}
\end{eqnarray}
where $I_{11}$ and $I_{12}$ can be written as the sum of the mean and covariance contributions:
\begin{eqnarray}
I_{11} & = & I_{11,\mathrm{mean}} + I_{11,\mathrm{cov}} \\
I_{12} & = & I_{12,\mathrm{mean}} + I_{12,\mathrm{cov}} .
\end{eqnarray}
We note that because the FIM is symmetric, $I_{12} = I_{21}$, and for large $n$, $I_{12}$ depends only on $\Delta s = |s_1 - s_2|$ and does not depend on the stimuli themselves. Moreover, because of the circular symmetry of the stimuli considered in this paper, $I_{11} = I_{22}$, and $I_{11}$ does not depend on the stimuli for large $n$. In the following, we will refer to the inverse of the asymptotic variance as encoding precision or, more commonly, as encoding accuracy.

{\bf Scaling of the asymptotic variance with the mean response gain $\alpha$ and the Fano factor:} In the expression for $I_{\mathrm{mean}}$ (Equation~\ref{dft_version}), $\mathbf{g}^i \propto \sqrt{\alpha}$ (this is because each of its entries, $\sigma_k^{-1}\partial f_k / \partial s_i$, scales as $\sqrt{\alpha}$ for Poisson-like noise). Similarly, the other $\mathbf{h}$ and $\mathbf{g}$ vectors also scale as $\sqrt{\alpha}$. Thus, each of the summands in Equation~\ref{dft_version} scales as $\alpha$, and therefore the whole expression scales as $\alpha$. It is easy to see that $I_{\mathrm{cov}}$ is independent of the gain (to see this, note that the entries of the $\mathbf{p}$ and $\mathbf{t}$ vectors in Equation~\ref{I_ij_cov_dft} are of the form shown in Equation~\ref{s_inv_s_partial} in which the gain term cancels). Hence from Equation~\ref{var_analytic}, we see that the asymptotic variance should have a weaker gain-dependence than $\alpha^{-1}$. This means that a doubling of the gain leads to a less than twofold increase in encoding precision. By a similar reasoning, it is easy to show that the asymptotic variance has a dependence on the Fano factor, FF, that is weaker than FF (because $I_{\mathrm{mean}}$ scales as $\mathrm{FF}^{-1}$ and $I_{\mathrm{cov}}$ is independent of FF). Again, this implies that halving the Fano factor should lead to a less than twofold increase in encoding precision. 

\subsection*{Additive noise: finite $n$}
If the noise is assumed to be additive, the covariance matrix $Q$ becomes stimulus-independent; hence, $I_{\mathrm{cov}} = 0$. Furthermore, the expression for $I_{\mathrm{mean}}$ (Equation~\ref{dft_version}) can be written in a more transparent form in this case, because it becomes possible to separate out the effect of mixing weights in the $\tilde{\mathbf{g}}$ and $\tilde{\mathbf{h}}$ vectors. Each of the terms on the right-hand side of Equation~\ref{dft_version} can be expressed as a sum over different Fourier modes. Considering the $I_{12,\mathrm{mean}}$ term first, we have:
\small
\begin{eqnarray}
I_{12,\mathrm{mean}} & = & \frac{1}{n} \bigg[ (\tilde{\mathbf{g}}^1)^T C (\tilde{\mathbf{g}}^2)^\ast + (\tilde{\mathbf{h}}^1)^T D (\tilde{\mathbf{g}}^2)^\ast + (\tilde{\mathbf{g}}^1)^T D (\tilde{\mathbf{h}}^2)^\ast + (\tilde{\mathbf{h}}^1)^T C (\tilde{\mathbf{h}}^2)^\ast  \bigg] \nonumber \\
& = & \frac{1}{n\sigma^2} \bigg[ 2 w_1 w_2 \sum_{k = 0}^{n-1} C_{kk}\tilde{\mathbf{m}}_k (\tilde{\mathbf{m}}_k \exp(-2\pi i k \delta / n))^\ast  +  (w_1^2 + w_2^2) \sum_{k = 0}^{n-1} D_{kk} \tilde{\mathbf{m}}_k(\tilde{\mathbf{m}}_k \exp(-2\pi i k \delta / n))^\ast  \bigg] \label{dft_shift_theorem} \nonumber \\
& = & \frac{1}{n\sigma^2} \bigg[ 2 w_1 w_2  \sum_{k = 0}^{n-1} |\tilde{\mathbf{m}}_k|^2 C_{kk} \cos(2\pi k \delta / n) +  (w_1^2 + w_2^2) \sum_{k = 0}^{n-1} |\tilde{\mathbf{m}}_k|^2 D_{kk} \cos(2\pi k \delta / n) \bigg] ,
\label{I12_mean_additive_dft}
\end{eqnarray}
\normalsize
where $\sigma^2$ is the variance of the additive noise, $\mathbf{m}$ is the vector of the derivatives of the von Mises tuning functions (Equation~\ref{vonmises}) with respect to $s$ and $\delta$ is an integer that expresses the difference between the two stimuli $s_1$ and $s_2$ in terms of the number of tuning function centers that separate them. Because of the circular nature of the stimuli, the Fisher information only depends on the angular distance between the stimuli. Hence, without loss of generality, we assume $s_1=0$ and thus the derivatives in $\mathbf{m}$ are all evaluated at $s=0$. In deriving Equation~\ref{dft_shift_theorem}, we used the fact $\mathbf{g}^i$ can be expressed as a scaled circular shift of $\mathbf{g}^j$ by $\delta$ and similarly as a circular shift of $\mathbf{h}^j$. In the Fourier domain, a circular shift corresponds to the multiplication of the $k$-th Fourier mode by $\exp(-2\pi i k \delta /n)$. In Equation~\ref{I12_mean_additive_dft}, we made use of the fact that $\tilde{\mathbf{m}}_k = \tilde{\mathbf{m}}_{N-k}$ (due to the circular nature of the stimulus space) and the identity $\exp(2\pi i k \delta /n ) + \exp(2\pi i (n-k) \delta /n ) = 2 \cos(2\pi k \delta /n)$. 

Similarly, for the $I_{11,\mathrm{mean}}$ term, we derive the following expression:
\begin{eqnarray}
I_{11,\mathrm{mean}} = \frac{1}{n\sigma^2} \bigg[ (w_1^2 + w_2^2) \sum_{k = 0}^{n-1} |\tilde{\mathbf{m}}_k|^2 C_{kk} +  2 w_1 w_2  \sum_{k = 0}^{n-1} |\tilde{\mathbf{m}}_k|^2 D_{kk} \bigg].
\label{I11_mean_additive_dft}
\end{eqnarray} 
In addition, $I_{22,\mathrm{mean}}=I_{11,\mathrm{mean}}$ and $I_{21,\mathrm{mean}}=I_{12,\mathrm{mean}}$ as usual. Finally, the asymptotic variance and correlation of estimates can be computed using Equations~\ref{var_analytic} and \ref{corr_analytic} and recalling that for additive noise $I_{\mathrm{cov}}=0$.

For the additive-noise model, if the neurons are assumed to be independent, we can also write $I_{11} = \sigma^{-2} \frac{\partial \mathbf{f}^T}{\partial s_1} \frac{\partial \mathbf{f}}{\partial s_1} = \sigma^{-2} \Big\|\frac{\partial \mathbf{f}}{\partial s_1} \Big\|^2$ and $I_{12} = \sigma^{-2} \frac{\partial \mathbf{f}^T}{\partial s_1} \frac{\partial \mathbf{f}}{\partial s_2}$ where $\sigma^2$ denotes the common noise variance. Plugging these in Equation~\ref{var_analytic}, we obtain the following proportionality relation for the asymptotic variance of the optimal estimator:
\begin{equation}
\mathrm{Var}(\hat{s}_1|\mathbf{s}) \propto \frac{\Big\|\frac{\partial \mathbf{f}}{\partial s_1} \Big\|^2}{\Big\|\frac{\partial \mathbf{f}}{\partial s_1} \Big\|^4 - \Big( \frac{\partial \mathbf{f}^T}{\partial s_1} \frac{\partial \mathbf{f}}{\partial s_2} \Big)^2} . \label{additive_ind_var}
\end{equation}
Inverting this proportionality yields Equation~\ref{additive_ind_prec} for the encoding precision, which is used below to provide a qualitative explanation for the stimulus-dependence of encoding accuracy.

{\bf Effect of heterogeneity in mixing weights on $I_{\mathrm{mean}}$:} Heterogeneity in the mixing weights can be accounted for by writing $\mathbf{g}^i = (\mathbf{w}+\delta \mathbf{w}) \circ \bar{\mathbf{g}}^i$ where we separated out the effect of mixing weights on $\mathbf{g}^i$ (note that we can do this for additive noise, but not for Poisson-like noise). In this expression, $\mathbf{w}$ denotes the mean weight that is constant across the neurons in the group and $\delta \mathbf{w}$ represents the stochastic component of the weight vector that is different from neuron to neuron. We denote the variance of the weights across the neurons by $\sigma_w^2$. We first note that heterogeneity in the mixing weights only affects the $C$ terms in $I_{11,\mathrm{mean}}$ in Equation~\ref{dft_version}, because the weight fluctuations are assumed to be uncorrelated across different groups and across the same group but for different stimuli. With this in mind, from Equation~\ref{dft_version}, we have: 
\begin{eqnarray}
\langle I_{11,\mathrm{mean}}^{\mathrm{het}} \rangle & = & I_{11,\mathrm{mean}}^{\mathrm{hom}} + \langle [\delta \mathbf{w}_1 \circ \bar{\mathbf{g}}^1]^T U C U^\ast [\delta \mathbf{w}_1 \circ \bar{\mathbf{g}}^1] \rangle + \langle [\delta \mathbf{w}_2 \circ \bar{\mathbf{h}}^1]^T U C U^\ast [\delta \mathbf{w}_2 \circ \bar{\mathbf{h}}^1] \rangle \nonumber \\
& = & I_{11,\mathrm{mean}}^{\mathrm{hom}} + \langle \sum_{i,j,k} [\delta \mathbf{w}_1 \circ \bar{\mathbf{g}}^1]_i^T U_{ij} C_{jj} U_{jk}^\ast [\delta \mathbf{w}_1 \circ \bar{\mathbf{g}}^1]_k \rangle + \langle \sum_{i,j,k} [\delta \mathbf{w}_2 \circ \bar{\mathbf{h}}^1]_i^T U_{ij} C_{jj} U_{jk}^\ast [\delta \mathbf{w}_2 \circ \bar{\mathbf{h}}^1]_k \rangle \nonumber \\
& = & I_{11,\mathrm{mean}}^{\mathrm{hom}} + \langle \sum_{i,j} [\delta \mathbf{w}_1 \circ \bar{\mathbf{g}}^1]_i^T U_{ij} C_{jj} U_{ji}^\ast [\delta \mathbf{w}_1 \circ \bar{\mathbf{g}}^1]_i \rangle + \langle \sum_{i,j} [\delta \mathbf{w}_2 \circ \bar{\mathbf{h}}^1]_i^T U_{ij} C_{jj} U_{ji}^\ast [\delta \mathbf{w}_2 \circ \bar{\mathbf{h}}^1]_i \rangle \nonumber \\
& = & I_{11,\mathrm{mean}}^{\mathrm{hom}} + \frac{1}{n} \big(\sum_{j} C_{jj} \big) \langle \sum_i (\delta \mathbf{w}_{1,i}\bar{\mathbf{g}_i}^1)^2 \rangle + \frac{1}{n} \big(\sum_{j} C_{jj} \big) \langle \sum_i (\delta \mathbf{w}_{2,i}\bar{\mathbf{h}_i}^1)^2 \rangle \nonumber \\
& = & I_{11,\mathrm{mean}}^{\mathrm{hom}} + \frac{\sigma_w^2}{n} \big(\sum_{j} C_{jj} \big) \big(\sum_i (\bar{\mathbf{g}_i}^1)^2 + (\bar{\mathbf{h}_i}^1)^2 \big) , 
\end{eqnarray}
where $I_{11,\mathrm{mean}}^{\mathrm{hom}}$ is the Fisher information for a homogeneous population derived above. In terms of $O(1)$ quantities, we can express $\langle I_{11,\mathrm{mean}}^{\mathrm{het}} \rangle$ as follows:
\begin{equation}
\langle I_{11,\mathrm{mean}}^{\mathrm{het}} \rangle  = I_{11,\mathrm{mean}}^{\mathrm{hom}} + n\sigma_w^2 M , \label{I_mean_hetero}
\end{equation}
where $M = (1/n^2) (\sum_{j} C_{jj}) (\sum_i (\bar{\mathbf{g}_i}^1)^2 + (\bar{\mathbf{h}_i}^1)^2) \sim O(1)$. Thus, the effect of heterogeneity is linear in the number of neurons per group. This scaling is the same as the corresponding scaling relationship derived previously in Ecker et al. (2011) for a neural population encoding a single stimulus. In Ecker et al. (2011), it is further shown that the same scaling holds for Poisson-like noise as well, but the variance $\sigma_w^2$ should, in that case, be interpreted as the variance of $\sqrt{\alpha}$, i.e. the square root of the gain rather than the variance of the gain itself.

\subsection*{Additive noise: large $n$}
In this section, we give a detailed analysis of the additive noise scenario in the limit of large $n$. We first derive explicit expressions for the $C$ and $D$ matrices (Equations~\ref{C_nn} and \ref{D_nn}) in the large $n$ limit. To do this, we first need to derive $\tilde{\mathbf{a}}$ and $\tilde{\mathbf{b}}$. It is convenient in this case to use indices ranging from $-(n-1)/2$ to $(n-1)/2$. For the derivations to be presented in this section, we adopt the following notation: $\omega = 2\pi / n$, $\lambda = \exp(-\omega / L)$ and $\tau = \exp(-i\omega j k)$. For $\tilde{\mathbf{a}}$, we have:
\begin{eqnarray}
\tilde{\mathbf{a}}_k & = & \sum_{j = -(n-1)/2}^{(n-1)/2} \big[ \delta_j + (1-\delta_j)c_0 \exp\big(-\frac{|\omega j|}{L}\big) \big]\exp(-i \omega j k) \nonumber \\
& = & 1 + c_0 \sum_{j \neq 0} \exp(-\frac{|\omega j |}{L}) \exp(-i \omega j k) \nonumber \\
& = & 1 + c_0 \sum_{j = 1}^{(n-1)/2} \exp(-\frac{\omega j}{L})\big[ \exp(-i \omega j k) + \exp(i \omega j k) \big] \nonumber \\
& = & 1 + c_0 \sum_{j = 1}^{(n-1)/2} (\lambda \tau)^j + (\lambda \tau^{-1})^j . 
\end{eqnarray}
We now take the sum of the two geometric series in the last equation. Denoting $\Gamma = \frac{n-1}{2} + 1$, we have:
\begin{equation}
\tilde{\mathbf{a}}_k = 1 + c_0 \bigg[ \frac{1-(\lambda \tau)^\Gamma}{1-\lambda \tau} + \frac{1-(\lambda \tau^{-1})^\Gamma}{1-\lambda \tau^{-1}} - 2 \bigg].
\end{equation}
After a little bit of algebra and rearranging, we obtain:
\begin{equation}
\tilde{\mathbf{a}}_k = 1 + c_0  \frac{\lambda^\Gamma (\lambda \cos((\Gamma-1) \omega k) - \cos(\Gamma \omega k)) + \lambda \cos(\omega k) -\lambda^2}{1- 2\lambda \cos(\omega k) + \lambda^2} ,  
\end{equation}
where we made repeated use of the identity $\exp(-i\theta) + \exp(i \theta) = 2\cos(\theta)$. We now consider the large $n$ value of the expression above by keeping only terms of leading order in $\omega = 2 \pi / n$. We recall that $\exp(x) = 1+ x + x^2 / 2 + O(x^3)$ and $\cos(x) = 1 - x^2 / 2 + O(x^4)$. The final result is: 
\begin{equation}
\tilde{\mathbf{a}}_k = 1 + \frac{c_0 n}{L \pi} \frac{1-\exp(-\pi/L)(-1)^k }{k^2 + L^{-2}}. \label{tilde_ak_large_n}
\end{equation}
This expression is the same as the one derived in Sompolinsky et al. (2001) for the same type of limited-range correlation structure. Proceeding similarly for $\tilde{\mathbf{b}}_k$, we find the following expression:
\begin{equation}
\tilde{\mathbf{b}}_k = \beta c_0 \Big[ 1 + \frac{n}{L \pi} \frac{1-\exp(-\pi/L)(-1)^k }{k^2 + L^{-2}} \Big]. \label{tilde_bk_large_n}
\end{equation}
Plugging these expressions in Equations~\ref{C_nn} and \ref{D_nn} for $C_{kk}$ and $D_{kk}$ and considering the large $n$ limit again, we arrive at the following large $n$ expressions for $C_{kk}$ and $D_{kk}$:
\begin{eqnarray}
C_{kk} =  \frac{1}{c_0 (1-\beta^2) \Phi_k} \label{large_n_C} \\
D_{kk} = -\frac{\beta}{c_0 (1-\beta^2) \Phi_k}, \label{large_n_D} 
\end{eqnarray}
where we denote:
\begin{equation}
\Phi_k = \frac{n}{L \pi} \frac{1-\exp(-\pi/L)(-1)^k }{k^2 + L^{-2}}. \label{big_phi_eq}
\end{equation}
By plugging these large $n$ expressions for $C_{kk}$ and $D_{kk}$ in Equations~\ref{I12_mean_additive_dft} and \ref{I11_mean_additive_dft}, we obtain the following large $n$ expressions for $I_{11,\mathrm{mean}}$ and $I_{12,\mathrm{mean}}$ for the case of additive noise: 
\begin{eqnarray}
I_{11,\mathrm{mean}} & = & \frac{w_1^2 + w_2^2 - 2 \beta w_1 w_2 }{n\sigma^2 c_0 (1-\beta^2) }  \sum_{k = 0}^{n-1} \frac{|\tilde{\mathbf{m}}_k|^2}{\Phi_k} \label{I11_mean_large_n}\\
I_{12,\mathrm{mean}} & = & \frac{ 2 w_1 w_2 - \beta (w_1^2 + w_2^2)}{n\sigma^2 c_0 (1-\beta^2)} \sum_{k = 0}^{n-1} \frac{|\tilde{\mathbf{m}}_k|^2 \cos(2\pi k \delta / n)}{ \Phi_k}.  \label{I12_mean_large_n}
\end{eqnarray}
As in Sompolinsky et al. (2001), it can be shown that $I_{11,\mathrm{mean}}$ and $I_{12,\mathrm{mean}}$ saturate to finite values when $c_0 \neq 0$ and diverge for $c_0 = 0$. To see this, write $\Psi_k = \Phi_k / n$. Similarly, write $\mathbf{\mu}_k = \tilde{\mathbf{m}}_k / n$. Then, considering $I_{11,\mathrm{mean}}$ as an example, we have :
\begin{equation}
I_{11,\mathrm{mean}} = \frac{w_1^2 + w_2^2 - 2 \beta w_1 w_2 }{\sigma^2 c_0 (1-\beta^2)} \sum_{k = 0}^{n-1} \frac{|\mathbf{\mu}_k|^2}{\Psi_k}.
\end{equation}
We note that $\Psi_k \sim O(k^{-2})$. Thus, if the power spectrum $|\mathbf{\mu}_k|^2$ decays sufficiently rapidly with $k$, e.g. $|\mathbf{\mu}_k|^2 \sim O(k^{-p})$ with $p>3$ (meaning that the tuning function derivatives are sufficiently smooth), the sum above remains $O(1)$ for $c_0 \neq 0$. An identical argument can be made for the sum in $I_{12,\mathrm{mean}}$ to show that $I_{11,\mathrm{mean}}$ and $I_{12,\mathrm{mean}}$ are both $O(1)$ for $c_0 \neq 0$ assuming sufficiently smooth tuning function derivatives.      

\subsubsection*{The effect of stimulus mixing on encoding accuracy} \label{liner_mixing_sec}

We can now ask what the effects of changing various parameters are on encoding accuracy in the case of additive noise. We first consider the effect of stimulus mixing. Assuming $w_1 = w$ and $w_2 = 1 - w$ and ignoring a common pre-factor which is always positive, we have (from Equations~\ref{I12_mean_additive_dft} and \ref{I11_mean_additive_dft}):
\begin{eqnarray}
I_{11,\mathrm{mean}} & \propto &   G \equiv (2w^2 - 2w) (X-Y) + X  \\
I_{12,\mathrm{mean}} & \propto &   H \equiv (2w^2 - 2w) (Z-T) + Z  \\
\mathrm{Var}(\hat{s}_1|\mathbf{s}) & \propto &  \frac{G}{G^2 - H^2},
\label{effect_of_stim_mixing_eq}
\end{eqnarray} 
where we use the following abbreviations:
\begin{eqnarray}
X & = & \sum_{k = 0}^{n-1} |\tilde{\mathbf{m}}_k|^2 C_{kk} \label{XYZT_eq_first} \\ 
Y & = & \sum_{k = 0}^{n-1} |\tilde{\mathbf{m}}_k|^2 D_{kk}                       \\
Z & = & \sum_{k = 0}^{n-1} |\tilde{\mathbf{m}}_k|^2 D_{kk} \cos(2\pi k \delta / n) \\
T & = & \sum_{k = 0}^{n-1} |\tilde{\mathbf{m}}_k|^2 C_{kk} \cos(2\pi k \delta / n). \label{XYZT_eq_last}
\end{eqnarray}

We now show that for the interval $0.5 < w < 1$, the variance (Equation~\ref{effect_of_stim_mixing_eq}) is a decreasing function of $w$. Therefore, stimulus mixing always reduces encoding accuracy and the stronger the mixing, the lower the encoding accuracy. For this purpose, it is sufficient to consider the sign of the derivative of Equation~\ref{effect_of_stim_mixing_eq} with respect to $2w^2 - 2w$ only, because using the chain rule, the derivative with respect to $w$ can be obtained by multiplying with $4w-2$ which is always positive in the interval $0.5 < w < 1$. Denoting the derivatives with respect to $2w^2 - 2w$ with a prime, we first observe the following: (i) $G^\prime =  X-Y \geq 0$; (ii) $H^\prime = Z-T$; (iii) $G^\prime > |H^\prime|$ and (iv) $G>0$. These are all easy to verify using the definitions in Equations~\ref{XYZT_eq_first}-\ref{XYZT_eq_last} and the large $n$ expressions for $C$ and $D$ (Equations~\ref{large_n_C}-\ref{large_n_D}).

Now, taking the derivative of the variance (Equation~\ref{effect_of_stim_mixing_eq}) and ignoring the denominator in the derivative which is always non-negative, we have:
\begin{eqnarray}
\big[\mathrm{Var}(\hat{s}_1|\mathbf{s})\big]^\prime & \propto & G^\prime ( G^2 - H^2 ) - ( G^2 - H^2 )^\prime G \nonumber \\
& = & G^\prime ( G^2 - H^2 ) - ( 2GG^\prime - 2HH^\prime ) G \nonumber \\
& = & -G^\prime (G^2 + H^2) + 2GHH^\prime \nonumber \\
& < & - |H^\prime|(G^2 + H^2) + 2 G |H| |H^\prime| \nonumber \\
& = & - |H^\prime| (G-|H|)^2 \leq 0 
\label{deriv_var_w}
\end{eqnarray}
Thus, the derivative is always negative and the variance is a decreasing function of $w$.

\subsubsection*{The effect of noise correlations on encoding accuracy} \label{linear_corr_sec}
In this section, we separately consider the effects of the three parameters determining the shape and the magnitude of noise correlations in the model: $c_0$, $\beta$, and $L$. Our strategy is to consider the derivative of the variance with respect to the parameter of interest and look at the sign of the derivative for different settings of the parameters. When the derivative is negative, the variance is a decreasing function of the parameter of interest; whereas a positive derivative means that the variance is an increasing function of the parameter of interest. We use the large $n$ expressions for the matrices $C$ and $D$ (thus for $I_{11,\mathrm{mean}}$, $I_{12,\mathrm{mean}}$ and the asymptotic variance as well) in the analyses to be presented below.

{\bf The effect of $c_0$:} We first consider the effect of changing $c_0$, the maximum correlation between any two neurons in the population. The analysis of the effect of $c_0$ is easier than the other two cases, because from the large $n$ expressions for $I_{11,\mathrm{mean}}$ and $I_{12,\mathrm{mean}}$ (Equations~\ref{I11_mean_large_n} and \ref{I12_mean_large_n}), we see that the effect of $c_0$ completely factorizes in these expressions and that they are both proportional to $1/c_0$. Thus, the variance is proportional to $c_0$. Hence, increasing $c_0$ is always harmful for encoding accuracy. This result is consistent with earlier results for homogeneous neural populations with additive noise encoding a single stimulus (Sompolinsky et al., 2001; Ecker et al., 2011). 

{\bf The effect of $\beta$:} We next consider the effect of $\beta$, the scaling parameter for across-group correlations. $\beta$ appears in a factorized form in $I_{11,\mathrm{mean}}$ and $I_{12,\mathrm{mean}}$ in the large $n$ limit (Equations~\ref{I11_mean_large_n} and \ref{I12_mean_large_n}) and therefore the derivative of the variance with respect to $\beta$ is relatively straightforward to compute. Figure~\ref{beta_L_encoding_accuracy}A shows the sign of $\partial \mathrm{Var}(\hat{s}_1|\mathbf{s}) / \partial \beta$ for different values of $w$, $\Delta s$ and $\beta$. 

{\bf The effect of $L$:} To investigate the effect of correlation length scale $L$ on encoding accuracy, we proceed similarly. In Figure~\ref{beta_L_encoding_accuracy}B, we plot the sign of $\partial \mathrm{Var}(\hat{s}_1|\mathbf{s}) / \partial L$ for different values of $w$, $\Delta s$ and $L$. This figure suggests that there is a critical value of $L$ around $L \approx 0.6$ below which the derivative is always positive, suggesting that it is beneficial to decrease $L$ in this regime. On the other hand, above the critical value of $L$, the derivative is always negative, suggesting that it is beneficial for encoding accuracy to increase $L$ in this regime. The critical value of $L$ decreases with the concentration parameter of the tuning functions $\gamma$ (Equation~\ref{vonmises}), but does not depend on any other parameters of the encoding model. This type of threshold-like behavior for the effect of correlation length scale $L$ is again consistent with a similar behavior reported in Sompolinsky et al. (2001).

\subsection*{Optimal linear estimator (OLE)} \label{ole_sec}
For the OLE, we first map the stimuli to Cartesian coordinates, $\mathbf{x} = \big[\cos(s_1) \, \sin(s_1) \, \cos(s_2) \, \sin(s_2) \big]$, and calculate the weight matrix $W$ that minimizes the mean squared error $\langle \| \mathbf{x} - \hat{\mathbf{x}} \|^2\rangle$ between the actual and estimated stimuli. The optimal weight matrix is given by: $W = Q_{\mathbf{rr}}^{-1} Q_{\mathbf{xr}}$ where $Q_{\mathbf{rr}}= \langle \mathbf{r}^T\mathbf{r}\rangle$ is the covariance matrix of the responses and $Q_{\mathbf{xr}} = \langle \mathbf{r}^T \mathbf{x}\rangle$ is the covariance between the stimuli and neural responses (here $\langle \cdot \rangle$ denotes an average over both stimuli $\mathbf{x}$ and noisy neural responses $\mathbf{r}$). These averages were computed over 8192 random samples of $\mathbf{x}$ (generated from uniformly distributed $s_1$ and $s_2$ values) and $\mathbf{r}$. The performance of the estimator was then measured by numerically estimating the mean squared error (MSE) for stimulus pairs of the form $(-s,s)$ with $s\in [0,\pi/2]$.  

\subsection*{The nonlinear mixing rule of Britten and Heuer (1999)}
For the nonlinear mixing model of Britten and Heuer (1999), the mean response, $f_k$, of a neuron to a pair of stimuli $(s_1,s_2)$ is given by Equation~\ref{brittenheuer99_eq}. We include a factor of 2 in the denominator in Equation~\ref{brittenheuer99_eq} to make the neural gains approximately independent of $\nu$. Britten and Heuer's original equation does not include this factor. The derivative of the mean response with respect to the stimulus $s_i$ is given by:
\begin{equation}
\frac{\partial f_k}{\partial s_i}  = a 2^{-1/\nu}\big[ f(s_1;\phi_k)^{\nu} + f(s_2;\phi_k)^{\nu} \big]^{\frac{1-\nu}{\nu}} f(s_i;\phi_k)^{\nu-1} f^\prime(s_i;\phi_k) ,
\end{equation} 
where $f$ denotes the von Mises tuning function (Equation~\ref{vonmises}) and $f^\prime$ its derivative. Given the mean responses and their derivatives with respect to each stimulus, expressions similar to Equations~\ref{dft_version} and \ref{I_ij_cov_dft} can be used to compute the Fisher information matrix for the nonlinear mixing model of Britten and Heuer (1999), taking into account that we assume an unsegregated population for this case (see {\it Results}).

\subsection*{A divisive form of stimulus mixing} \label{divisive_sec}
In the divisively normalized stimulus mixing model, the response of a neuron in the first group is described by Equation~\ref{div_norm_mean_resp_eq} below. Responses of neurons in the second group are similar, but with the roles of $s_1$ and $s_2$ reversed in the right hand side. In Equation~\ref{div_norm_mean_resp_eq}, $f(s;\phi)$ is the von Mises tuning function defined in Equation~\ref{vonmises} and the weighting profile $w(\phi_k, \phi_{k^\prime})$ is a normalized von Mises function given by:
\begin{equation}
w(\phi_k, \phi_{k^\prime}) = \frac{\exp(\gamma_w [\cos(\phi_k - \phi_{k^\prime}) - 1])}{\sum_{k^\prime} \exp(\gamma_w [\cos(\phi_k - \phi_{k^\prime}) - 1])}.
\end{equation}

The derivatives of the mean response $f_k$ with respect to $s_1$ and $s_2$ are given by:
\begin{eqnarray}
\frac{\partial f_k}{\partial s_1} & = & \frac{2 f(s_1; \phi_k) f^\prime(s_1; \phi_k)}{\Delta + k_w \sum_{k^\prime} w(\phi_k, \phi_{k^\prime}) f(s_2; \phi_{k^\prime})^2} \label{divisive_norm_deriv_eq1}\\
\frac{\partial f_k}{\partial s_2} & = &  -\frac{2 f(s_1; \phi_k)^2 \sum_{k^\prime} w(\phi_k,\phi_{k^\prime}) f^\prime(s_2;\phi_{k^\prime}) f(s_2; \phi_{k^\prime})}{\big[ \Delta + k_w \sum_{k^\prime} w(\phi_k, \phi_{k^\prime}) f(s_2; \phi_{k^\prime})^2 \big]^2},  \label{divisive_norm_deriv_eq2}
\end{eqnarray}
where $f^\prime$ denotes the derivative of the von Mises tuning function.

Equations~\ref{dft_version} and \ref{I_ij_cov_dft} for the Fisher information matrix are still valid for the divisively normalized mixing model with the difference that the mean responses of neurons and their derivatives with respect to the two stimuli are computed according to Equations~\ref{div_norm_mean_resp_eq}, \ref{divisive_norm_deriv_eq1} and \ref{divisive_norm_deriv_eq2} respectively.

Figure~\ref{divisive_all_fig} shows the effect of varying the divisive normalization scaling factor $k_w$ and the across-group neural correlations $\beta$ on encoding accuracy for both the divisively normalized mixing model and a model where the off-diagonal terms in the Fisher information matrix ($I_{ij}$ with $i \neq j$) were set to zero, but the diagonal terms were the same as in the divisively normalized mixing model. As explained in the \textit{Results} section, this latter model eliminates stimulus mixing, but preserves the neuron-by-neuron response gains in the divisively normalized model. 

\subsection*{Stimulus mixing is not always harmful for encoding accuracy} \label{optimal_mixing_sec}
We first analyze the general stimulus mixing problem with a two-dimensional toy model. We imagine two ``neurons'' mixing the two stimuli $s_1$, $s_2$ according to $f_1(s_1,s_2)$ and $f_2(s_1,s_2)$ respectively. The responses of the two neurons are given by $r_1 = f_1(s_1,s_2) + \varepsilon_1$ and $r_2 = f_2(s_1,s_2) + \varepsilon_2$, where $\varepsilon_1$ and $\varepsilon_2$ are Gaussian random variables with variance $\sigma^2$ and correlation $\rho$. Thus, in the following analysis, we assume stimulus-independent additive noise. We denote the Jacobian matrix for the mean responses of the neurons by $\mathbf{J}$:
\begin{equation}
\mathbf{J} = \large \begin{bmatrix}
       \frac{\partial f_1}{\partial s_1} & \frac{\partial f_1}{\partial s_2} \\[0.5em]
       \frac{\partial f_2}{\partial s_1} & \frac{\partial f_2}{\partial s_2} 
\end{bmatrix}.
\label{jacobian_responses}
\end{equation}
As explained in the {\it Results} section, one can think of $\mathbf{J}$ as a mixing matrix describing the sensitivity of each neuron to changes in each stimulus. The Fisher information matrix is given by $I_F = \mathbf{J}^T \mathbf{\Sigma}^{-1}\mathbf{J}$ where $\mathbf{\Sigma}$ is the covariance matrix of the response noise. The inverse of $I_F$ gives the asymptotic covariance matrix of the maximum likelihood estimator. To find the optimal mixing matrix $\mathbf{J}$, we minimize the trace of $I_F^{-1}$, i.e. $\mathrm{Tr}[I_F^{-1}] = \mathrm{Tr}[\mathbf{J}^{-1} \mathbf{\Sigma} \mathbf{J}^{-T}]$ with respect to $\mathbf{J}$, subject to the constraint that the sum of the squares of the derivatives in $\mathbf{J}$ be a finite constant $K$, i.e. $\mathrm{Tr}[\mathbf{J}^T \mathbf{J}] = K$.

We find the optimal $\mathbf{J}$ by the method of Lagrange multipliers. The objective function is given by:
\begin{equation}
\mathcal{L} = \mathrm{Tr}[\mathbf{J}^{-1} \mathbf{\Sigma} \mathbf{J}^{-T}] + \lambda (\mathrm{Tr}[\mathbf{J}^T \mathbf{J}] - K)
\end{equation}
and the required derivatives are as follows: 
\begin{eqnarray}
\frac{\partial \mathcal{L}}{\partial \mathbf{J}} & = &  -2 \mathbf{J}^{-T} \mathbf{J}^{-1}\mathbf{\Sigma} \mathbf{J}^{-T} + 2\lambda \mathbf{J} \\
\frac{\partial \mathcal{L}}{\partial \lambda} & = & \mathrm{Tr}[\mathbf{J}^T \mathbf{J}] - K .
\end{eqnarray}
Setting these to zero and rearranging, we get the following equations:
\begin{eqnarray}
\mathbf{J}\mathbf{J}^T\mathbf{J}\mathbf{J}^T & = & \mathbf{Y}^2 = \lambda^{-1}\mathbf{\Sigma} \label{lagrange_first}\\ 
\mathrm{Tr}[\mathbf{J}^T \mathbf{J}] & = & \mathrm{Tr}[\mathbf{Y}] = K , \label{lagrange_second}
\end{eqnarray} 
where we denote $\mathbf{J}\mathbf{J}^T$ by $\mathbf{Y}$. By taking the eigendecomposition of the right-hand side of Equation~\ref{lagrange_first}, we obtain:
\begin{equation}
\mathbf{Y} = \sqrt{\lambda^{-1}}\mathbf{P}\mathrm{diag}(\sqrt{\lambda_1}, \sqrt{\lambda_2}) \mathbf{P}^{-1}, \label{Y_eigs}
\end{equation}  
where $\mathbf{P}$ is the matrix of eigenvectors of $\mathbf{\Sigma}$ and $\lambda_1$ and $\lambda_2$ are its eigenvalues. Because the trace of a matrix is equal to the sum of its eigenvalues, from Equation~\ref{lagrange_second}, we get $\sqrt{\frac{\lambda_1}{\lambda}} + \sqrt{\frac{\lambda_2}{\lambda}} = K$ or $\sqrt{\lambda^{-1}} = K/(\sqrt{\lambda_1} + \sqrt{\lambda_2})$. Plugging this in Equation~\ref{Y_eigs}:
\begin{equation}
\mathbf{Y} = \frac{K}{\sqrt{\lambda_1}+\sqrt{\lambda_2}}\mathbf{P}\mathrm{diag}(\sqrt{\lambda_1}, \sqrt{\lambda_2}) \mathbf{P}^{-1}.
\label{Y_final}
\end{equation}
The matrix of eigenvectors and the eigenvalues of the response covariance matrix $\mathbf{\Sigma}$ are given by:
\begin{equation}
\mathbf{P} = \large \rho \begin{bmatrix}
       1 & 1 \\[0.5em]
       -1 & 1 
\end{bmatrix}
\end{equation}
\begin{equation}
\lambda_1 = \sigma^2(1-\rho), \qquad \lambda_2 = \sigma^2(1+\rho) .
\end{equation}
Plugging these expressions in Equation~\ref{Y_final} and simplifying, we get:
\begin{equation}
\mathbf{Y} = \frac{K}{2} \begin{bmatrix}
       1 & \frac{1-\sqrt{1-\rho^2}}{\rho} \\[0.5em]
       \frac{1-\sqrt{1-\rho^2}}{\rho} & 1 
\end{bmatrix}.
\label{Y_final_final}
\end{equation}
Now, the $\mathbf{Y}$ matrix can be written as: 
\begin{equation}
\mathbf{Y} = \begin{bmatrix}
       \nabla f_1 \\[0.5em]
       \nabla f_2 
\end{bmatrix} \begin{bmatrix}
       \nabla f_1^T \; \nabla f_2^T 
\end{bmatrix} = \begin{bmatrix}
       \| \nabla f_1 \|^2 & \nabla f_1 \cdot \nabla f_2 \\[0.5em]
       \nabla f_1 \cdot \nabla f_2 & \| \nabla f_2 \|^2
\end{bmatrix},
\end{equation}
where $\nabla f_1$ and $\nabla f_2$ denote the gradients of $f_1$ and $f_2$ respectively. The cosine of the angle $\theta^\ast$ between these two gradients is given by:
\begin{equation}
\cos \theta^\ast  = \frac{\nabla f_1 \cdot \nabla f_2}{\| \nabla f_1 \| \| \nabla f_2 \|} = \frac{1-\sqrt{1-\rho^2}}{\rho}. \label{cos_theta}
\end{equation}
The optimal solution is thus to set the gradients to have equal norm ($\sqrt{K/2}$) and the angle between them to $\theta^\ast$ with $\cos \theta^\ast$ as given in Equation~\ref{cos_theta}.  Because $\cos$ is an even function, $\theta^\ast$ and $-\theta^\ast$ are both solutions. Figure~\ref{optimal_mixing_fig}A plots the positive solution as a function of $\rho$.

The solution of the two-dimensional toy model can be readily generalized to models with more than two neurons under certain assumptions. Consider, for instance, two groups of neurons with responses given respectively by:
\begin{eqnarray}
f_{1,k}(s_1,s_2) 		& = & w_{1} f(s_1;\phi_k) + w_{2} f(s_2;\phi_k) \label{linear_mixing_eq_first}\\
f_{2,k^\prime}(s_1,s_2) & = & w_{3} f(s_1;\phi_{k^\prime}) + w_{4} f(s_2;\phi_{k^\prime}).
\label{linear_mixing_eq}
\end{eqnarray} 
This model is very similar to the linear mixing model analyzed in detail in this paper (see {\it Results}), with the only difference being that the restriction for the weights to be positive and symmetric between the groups is now lifted. Assuming $s_1 \approx s_2 = s$ and an additive noise model, the problem of finding the optimal weights $w_1$, $w_2$, $w_3$ and $w_4$ subject to a total power constraint on the derivatives, i.e.:
\begin{equation}
(w_1^2 + w_2^2 + w_3^2 + w_4^2) \sum_{k} f^\prime(s;\phi_k)^2 = P 
\end{equation}  
can be directly translated into the two-dimensional problem with the following transformation: 
\begin{eqnarray}
\mathbf{J} & = & \large \begin{bmatrix}
        w_1 & w_2 \\[0.5em]
        w_3 & w_4 
\end{bmatrix} \\
\mathbf{\Sigma}^{-1} & = & \begin{bmatrix}
       \mathbf{f}^\prime & \mathbf{0} \\[0.5em]
       \mathbf{0} & \mathbf{f}^\prime 
\end{bmatrix} Q^{-1} \begin{bmatrix}
       \mathbf{f}^\prime & \mathbf{0} \\[0.5em]
       \mathbf{0} & \mathbf{f}^\prime 
\end{bmatrix}^T \\
K & = & \frac{P}{\sum_{k} f^\prime(s;\phi_k)^2}, \label{resource_K}
\end{eqnarray}
where $\mathbf{f}^\prime$ is a row vector of the derivatives $df(s;\phi_k)/ds$, $\mathbf{0}$ is a row vector of zeros and $Q$ is the covariance structure of the neurons (e.g., $Q = SRS$ with $R$ as given in Equation~\ref{R_block_matrix}). In particular, for uncorrelated neurons (diagonal $Q$), we find that the optimal solution is to set the weight vectors $\big[ w_1, w_2\big]$ and $\big[ w_3, w_4\big]$ such that they have equal norm and are orthogonal to each other. 

For the more general case of $n$ neurons encoding two stimuli, as far as we know, there is no closed-form solution for the optimal mixing matrix subject to a constraint on the total power of the derivatives. We thus solve this more general problem numerically. Figure~\ref{three_solutions_fig} shows three distinct solutions for $n=16$ with both a diagonal covariance matrix (Figure~\ref{three_solutions_fig}A) and a limited-range correlation structure (Figure~\ref{three_solutions_fig}B) as in Equation~\ref{A_matrix} with $c_0 = 0.3$ and $L=2$.      

Similarly, for the linear encoding model (Equations~\ref{linear_mixing_eq_first}-\ref{linear_mixing_eq}), when $s_1 \not\approx s_2$, it does not seem possible to reduce the problem of finding the optimal mixing weights $w_1$, $w_2$, $w_3$ and $w_4$ to our two-dimensional toy problem. Thus, we solve this optimization problem numerically as well. Because $I_F^{-1}$ depends on $\Delta s$, we minimize the following objective function in this case: 
\begin{equation}
\langle \mathrm{Tr}[I_F^{-1}(\Delta s)] \rangle \propto \int \mathrm{Tr}[I_F^{-1}(\Delta s)] d\Delta s ,
\end{equation}
where the integral is approximated by the average of $ \mathrm{Tr}[I_F^{-1}(\Delta s)]$ over 21 $\Delta s$ values uniformly spaced in the interval $[0,\pi]$. The total resource $K$ is assumed to be equal to the number of neurons and the noise is assumed to be additive (and the noise variance identical for all neurons) both in the arbitrary encoding model and in the linear encoding model. All numerical optimization problems were solved using the genetic algorithm routines in Matlab's Global Optimization Toolbox.

\subsection*{Parameter values for the results reported in the figures} \label{parameter_vals}

For our main results, reported in Figures~1, 2, 4, 7, 8, 9, 10, we use a Poisson-like noise model and a limited-range noise correlation structure with parameters $c_0=0.3$ and $L=2$, which is consistent with the small but broad noise correlations typically observed in the visual cortical areas (Cohen and Kohn, 2011). For the tuning function parameters, we again use values broadly consistent with response properties of neurons in the visual cortex: $\alpha = 20$, $\gamma = 2$, $\eta = 0$ (Ecker et al., 2011). For convenience, we assume tuning function centers that are uniformly spaced between $0$ and $2\pi$. 

In Figures~3, 11, 12, we use an additive Gaussian noise model. The additive-noise assumption is needed to establish the analytic results regarding the effects of varying the parameters of the encoding model on encoding accuracy, as well as for the solution of the optimal mixing model presented at the end of the \textit{Results} section. 

For the number of neurons, we used as large a number of neurons as possible. Specifically, in Figures~1, 2, 4, 6, 10, we use $n=4096$ (number of neurons per group). In Figure~9, since there is only a single, unsegregated population, we use $n=8192$. Due to computational costs, we had to use a smaller number of neurons per group in Figures~7 and 8: in Figure~7, $n=1024$ (note this means that the total number of neurons is $2048$ for set size $N=2$ and 6144 for $N=6$) and in Figure~8, $n=1024$.

Other parameter values specific to each figure are as follows: in Figure~9, $a=1$, $b=0$ (parameters of the nonlinear mixing model of Britten and Heuer). In Figure~10, $\Delta = 10$, $\gamma_w = 2$ (divisive normalization parameters). In Figure~12, for the linear encoding model, tuning function parameters are the same as those reported above for Figure 1.

\section*{Results}
\subsection*{Linear mixing}
We consider a population of neurons encoding a pair of stimuli $(s_1,s_2)$ where the mean responses of the neurons are expressed as a weighted average of their responses to the individual stimuli: 
\begin{equation}
f_k(s_1,s_2) = w_{1} f(s_1;\phi_k) + w_{2} f(s_2;\phi_k)  .
\label{linear_mixing_mean}
\end{equation}
Here $w_{1}$ and $w_{2}$ are the mixing weights and $\phi_k$ is the prefered stimulus of neuron $k$. We call this type of mixed selectivity \emph{linear mixing}, although it should be noted that the mean response $f_k(s_1,s_2)$ is linear only in the individual responses and not in the stimuli themselves, therefore linear mixing in this sense is different from what is referred to as linear mixing in Rigotti et al. (2013). Neural responses of this type have been observed throughout the cortex (Recanzone et al., 1997; Britten and Heuer, 1999; Zoccolan et al., 2005; Zoccolan et al., 2007; Beck and Kastner, 2007; Busse et al., 2009; MacEvoy et al., 2009; MacEvoy and Epstein, 2009; Nandy et al., 2013). It is, thus, of considerable interest to understand the information encoding consequences of this type of mixed selectivity. For our analysis, we separate the neurons into two groups such that neurons in the first group have a larger weight for the first stimulus and neurons in the second group have a larger weight for the second stimulus. For simplicity, we assume symmetric weights for the two groups: i.e. if the mixing weights associated with $s_1$ and $s_2$ are $w_1$ and $w_2$ respectively for the first group (with $w_1 \geq w_2$), they are $w_2$ and $w_1$ for the second group. We initially consider the case where all neurons within the same group have the same weights for the two stimuli, but later consider the effects of heterogeneity in mixing weights. 

To model variability in neural responses, unless otherwise noted, we use a biologically realistic, Poisson-like noise model, where the variance of the noise is equal to the mean response ({\it Materials and Methods}). We assume that neural variability is correlated across the population. Specifically, within-group correlations between neurons decay exponentially with the distance between their prefered stimuli, consistent with experimental measurements of noise correlations throughout the visual cortex (Cohen and Kohn, 2011):
\begin{equation}
R_{kl} = \delta_{kl} + (1-\delta_{kl})c_0\exp\bigg(-\frac{|\phi_k - \phi_l|}{L}\bigg),
\label{R_matrix}
\end{equation}
where $\delta$ is the Kronecker delta function. Across-group correlations are simply scaled versions of within-group correlations:
\begin{equation}
R_{kl^\prime} = \beta c_0 \exp\bigg(-\frac{|\phi_k - \phi_{l^\prime}|}{L}\bigg),
\label{R_prime_matrix}
\end{equation}
where $0 \leq \beta \leq 1$ represents the scaling factor. In this paper, we only consider stimuli defined over circular spaces due to their conceptual simplicity and analytical tractability. Stimuli defined over bounded spaces would introduce edge effects where stimuli toward the edges are encoded with less accuracy than stimuli toward the center. This can be explained entirely by a decrease in the effective number of neurons covering the stimuli toward the edges and hence is uninteresting for our considerations. We do not expect any of our main results concerning stimulus mixing to depend on the choice of a circular rather than a bounded stimulus space.

\subsection*{Consequences of linear mixing}
We derived a mathematical expression for the Fisher information matrix (FIM) of the encoding model described above. The main interest in deriving the FIM comes from the fact that, by the Cram\'{e}r-Rao bound, the inverse of the FIM provides a lower bound on the covariance matrix of any unbiased estimator of the stimuli and expresses the asymptotic covariance matrix of the maximum-likelihood estimator. From the inverse of the FIM, we obtained expressions for the asymptotic variance of the optimal estimates of $s_1$ and $s_2$ and the correlation between the estimates ({\it Materials and Methods}). 

\begin{figure}
\centerline{\includegraphics[width=.99\textwidth,trim=0mm 0mm 0mm 0mm,clip]{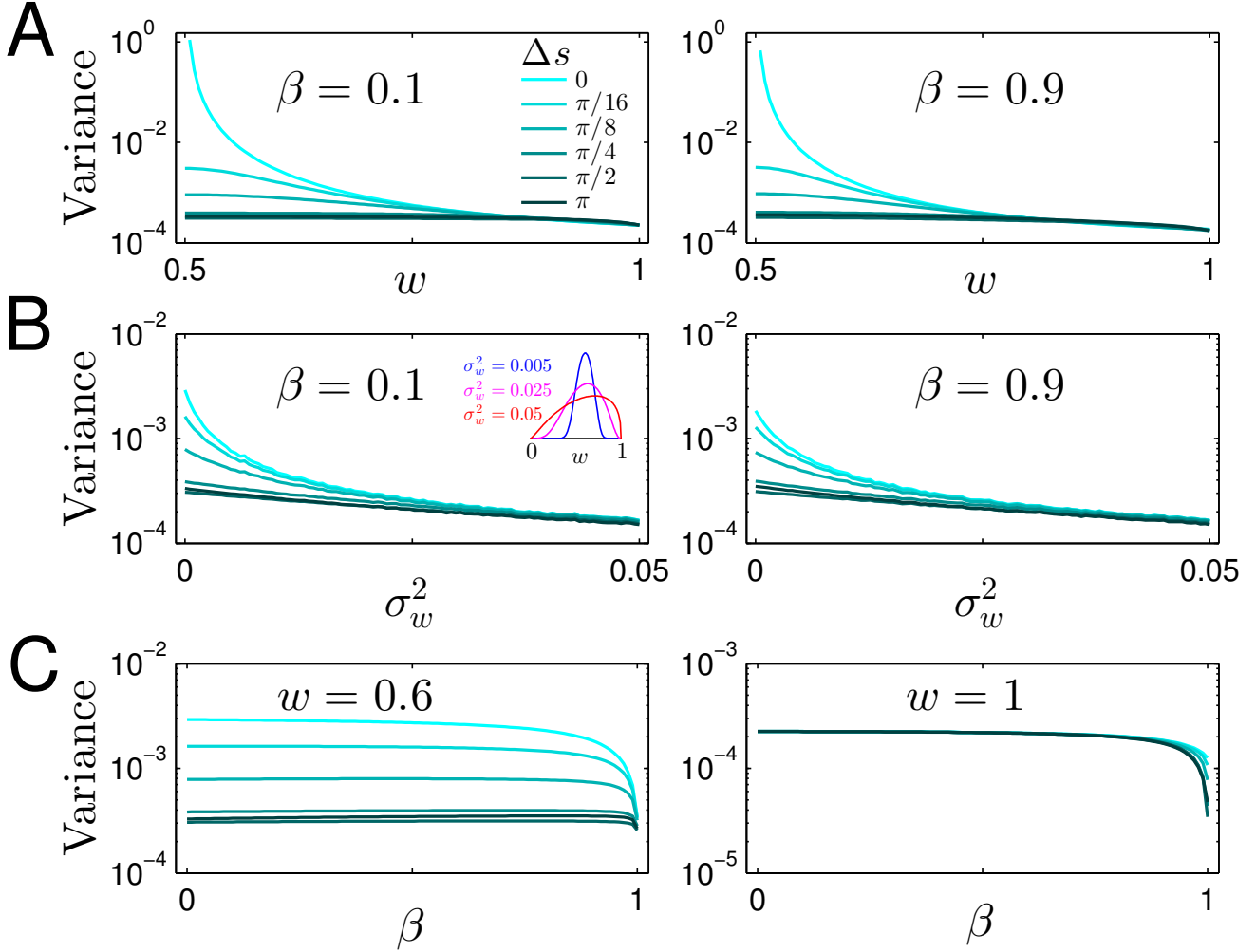}}
\caption{Analysis of the linear mixing model. (A) The asymptotic variance of the optimal estimator, $\mathrm{Var}(\hat{s}_1|\mathbf{s})$, as a function of the mixing weight $w$. The left panel shows the optimal variance for a small $\beta$ value ($\beta=0.1$) and the right panel for a large $\beta$ value ($\beta=0.9$). (B) The effect of heterogeneity in mixing weights on encoding accuracy. The weights are drawn from a beta distribution with mean $0.6$ or $0.4$, and variance $\sigma_w^2$. The two panels show the optimal variance for two different $\beta$ values. To aid the interpretation of $\sigma_w^2$ values, the inset in the left panel shows the weight distributions for three different $\sigma_w^2$ values. (C) The asymptotic variance of the optimal estimator as a function of $\beta$. The left panel shows the optimal variance for a strong stimulus mixing regime ($w=0.6$) and the right panel for the no stimulus mixing regime ($w=1$). Different curves in each panel correspond to six different $\Delta s=|s_1 - s_2|$ values {\color{black}indicated in the inset in (A)}, with lighter colors corresponding to smaller $\Delta s$ values. Other parameter values are listed in {\it Materials and Methods}.}\label{w_beta_hetero_fig}
\end{figure}

We then asked how changes in different parameters affect encoding accuracy, i.e. the inverse of the asymptotic variance of the estimates. Considering the effect of stimulus mixing first and assuming $w_1 = w$ and $w_2 = 1 - w$ with $0.5 \leq w \leq 1$, we find that increased stimulus mixing (i.e., decreasing $w$) reduces encoding accuracy and that these reductions can be substantial (Figure~\ref{w_beta_hetero_fig}A). The harmful effect of stimulus mixing for encoding accuracy depends on the similarity between the two stimuli (Figure~\ref{w_beta_hetero_fig}A), being more severe for more similar stimuli (smaller $\Delta s = |s_1 - s_2|$). For some stimulus pairs, {\color{black}increased stimulus mixing can cause several orders of magnitude reductions in encoding accuracy} (Figure~\ref{w_beta_hetero_fig}A). It is easy to see that the total response across the whole population is independent of the mixing weight $w$. Therefore, the reduction in encoding accuracy with increased stimulus mixing is due entirely to stimulus mixing itself, rather than any reduction in the overall response level.

We next analyzed the effect of heterogeneity in mixing weights by assuming that the weights of different neurons are drawn from a distribution with mean $w$ or $1-w$ and variance $\sigma_w^2$ ({\it Materials and Methods}). Such heterogeneity in the mixing weights partially alleviates the harmful effects of stimulus mixing (Figure~\ref{w_beta_hetero_fig}B). Increasing the across-group neural correlations, i.e. increasing $\beta$, can also counteract the effects of stimulus mixing under certain parameter regimes (Figure~\ref{w_beta_hetero_fig}C). 

\begin{figure}
\centering
\includegraphics[width=.99\textwidth,trim=0mm 0.5mm 0mm 0mm,clip]{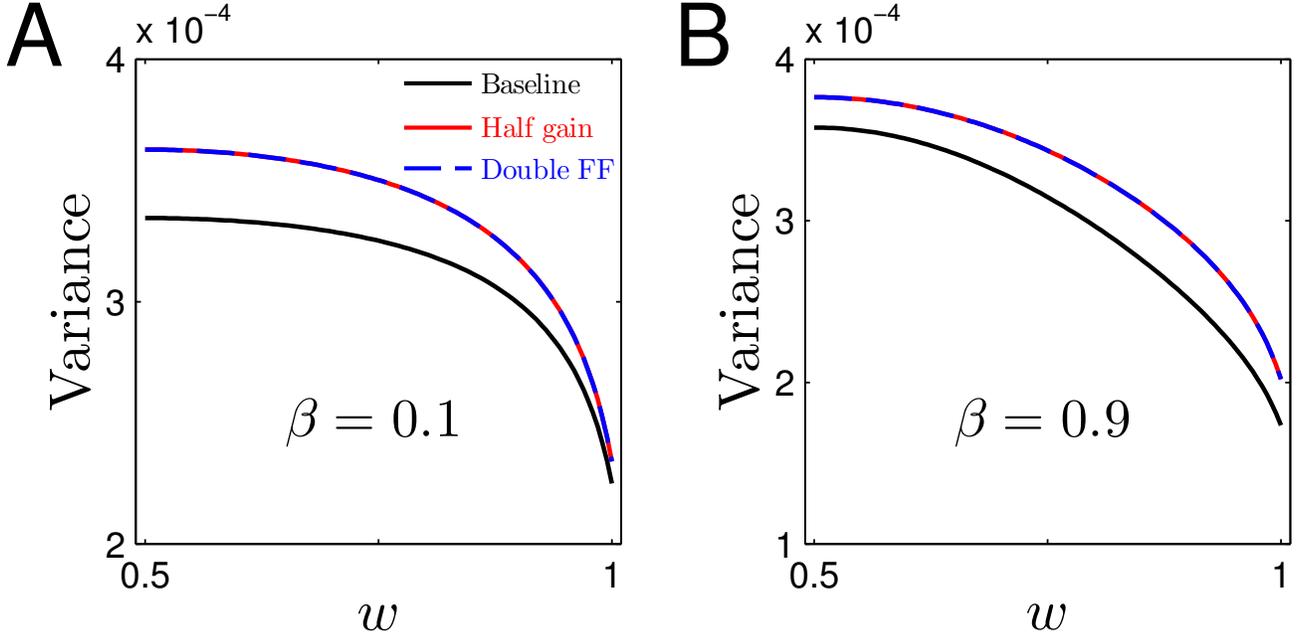} 
\caption{Comparison of the effect of stimulus mixing on the asymptotic variance with the effects of halving the mean response gain $\alpha$, or doubling the Fano factor, FF. (A) Weak across-group correlations ($\beta=0.1$). (B) Strong across-group correlations ($\beta=0.9$). The asymptotic variance has the same scaling with $\alpha$ as it does with $\mathrm{FF}^{-1}$. The results shown in this figure are for $\Delta s = \pi$, for which the effect of stimulus mixing is among the weakest. For smaller $\Delta s$ values, stimulus mixing generally has a much larger effect. The parameter values for the baseline results (shown in black) are the same as those reported for Figure~\ref{w_beta_hetero_fig} (see {\it Materials and Methods}).}
\label{gain_ff_w_comparison}
\end{figure}

{\bf Effects of stimulus mixing, heterogeneity in mixing weights and across-group neural correlations on encoding accuracy:} The presence of Poisson-like noise makes an analytic quantification of the effects of stimulus mixing, heterogeneity in mixing weights and across-group neural correlations on the asymptotic variance difficult. However, for the parameter regimes reported in Figure~\ref{w_beta_hetero_fig}, we numerically checked and confirmed the following: (i) increasing stimulus mixing always increases the asymptotic variance (Figure~\ref{w_beta_hetero_fig}A); (ii) increasing the heterogeneity of mixing weights generally reduces the asymptotic variance, except for a small number of cases where this pattern is reversed for very close $\sigma_w^2$ values due to stochasticity in sampling (Figure~\ref{w_beta_hetero_fig}B); (iii) for all $\Delta s$, the increase in variance caused by halving $w_1=w$ from $1$ to $0.5$ is always greater than the increase in variance caused by halving the mean response gain or doubling the Fano factor. Figure~\ref{gain_ff_w_comparison} compares the effect of stimulus mixing on the asymptotic variance with the effects of halving the mean response gain or doubling the Fano factor (FF). The results shown in Figure~\ref{gain_ff_w_comparison} are for $\Delta s = \pi$, for which the effect of stimulus mixing is among the weakest. For smaller $\Delta s$ values, stimulus mixing generally has a much larger effect on the variance than the effect of changing the gain or the Fano factor. This result could explain why attention acts primarily by stimulus selection, which in our model corresponds to changing the mixing weights, rather than simply through noise reduction or gain increase (Pestilli et al., 2011), because the former mechanism typically leads to a much larger improvement in encoding accuracy than the latter mechanisms (see {\it Discussion}).

\begin{figure}
\centering
\includegraphics[width=.99\textwidth,trim=0mm 0mm 0mm 0mm,clip]{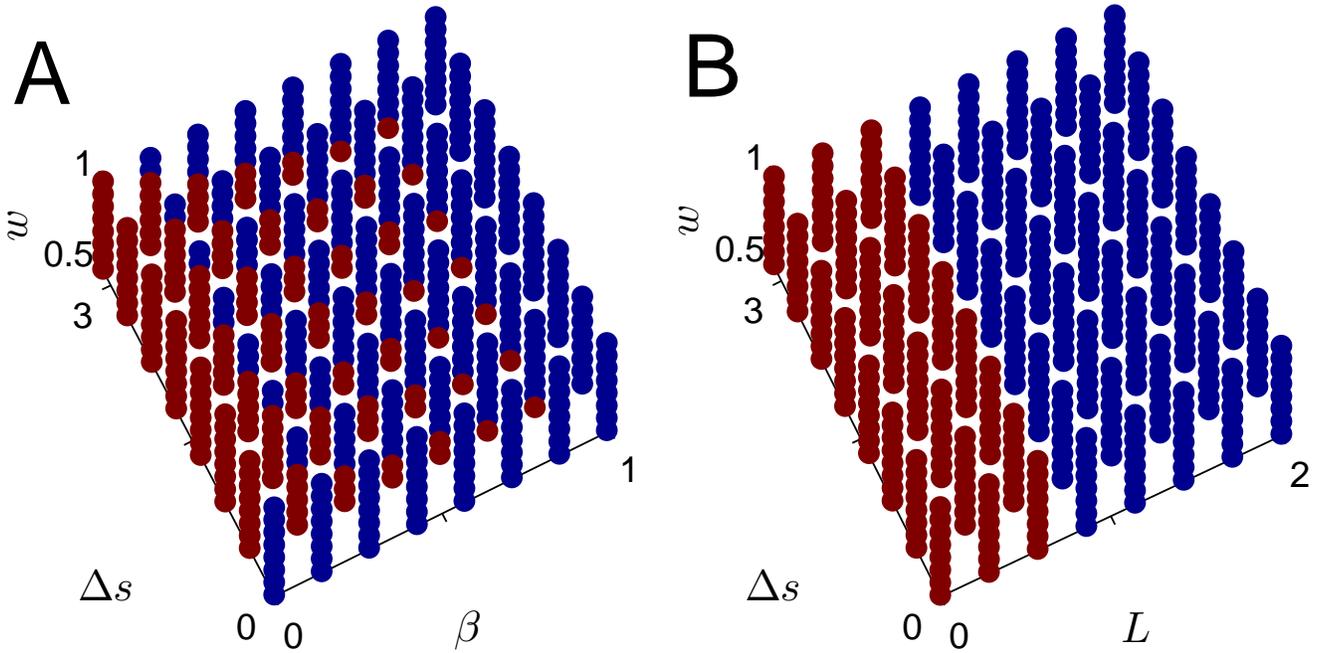} 
\caption{(A) The sign of $\frac{d\mathrm{Var}(\hat{s}_1|\mathbf{s})}{d\beta}$ as a function of $\beta$, $\Delta s$ and $w$. Red dots indicate parameter combinations for which the derivative is positive, blue dots indicate negative derivatives. $w$ is varied from $0.51$ to $0.99$, $\beta$ is varied from $0$ to $0.99$ and $\Delta s$ is varied from $0$ to $\pi$, all on a linear scale. $L=2$ for the results shown in this plot. (B) The sign of $\frac{d\mathrm{Var}(\hat{s}_1|\mathbf{s})}{dL}$ as a function of $L$, $\Delta s$ and $w$. $w$ is varied from $0.51$ to $0.99$, $L$ is varied from $10^{-6}$ to $2$ and $\Delta s$ is varied from $0$ to $\pi$. The results shown in this figure are for the additive noise model.}
\label{beta_L_encoding_accuracy}
\end{figure}

\textbf{Analytical results for an additive noise model:} To develop a better understanding of the effects of changing the parameters of the encoding model on encoding accuracy, we supplement the numerical results for Poisson-like noise with analytical results for a simpler additive noise model. In the additive noise model, the noise variance, rather than being equal to the mean response as in the Poisson-like noise model, is assumed to be the same for all neurons independent of their mean responses. In {\it Materials and Methods}, for the additive noise model and in the limit of a large number of neurons, we mathematically show that increased stimulus mixing always reduces encoding accuracy, increased heterogeneity in mixing weights always improves encoding accuracy, whereas the conditions under which increased neural correlations between the groups, i.e. increasing $\beta$, improves encoding accuracy are slightly more complicated. In Figure~\ref{beta_L_encoding_accuracy}A, we plot the sign of $\partial \mathrm{Var}(\hat{s}_1|\mathbf{s}) / \partial \beta$ for different values of $w$, $\Delta s$ and $\beta$. A positive sign means that the asymptotic estimation variance is an increasing function of $\beta$ (i.e. it is harmful to increase $\beta$), whereas a negative sign means that the asymptotic estimation variance is a decreasing function of the parameter. This figure shows that the derivative is always negative for very small values of $\Delta s$, suggesting that it is beneficial (in terms of encoding accuracy) to increase the across-group correlations in this case. For other values of $\Delta s$, the sign depends on $\beta$ and $w$, being more likely to be negative for larger $\beta$ and larger $w$ values. A detailed analysis of the effects of changing the other parameters of the encoding model under the additive noise assumption can also be found in the {\it Materials and Methods} section. In summary, this analysis shows that increasing the maximum neural correlation, $c_0$, is always harmful for encoding accuracy, whereas for the correlation length scale $L$, there is a critical threshold below which it is always harmful to increase $L$ and above which it is always beneficial to increase $L$ (Figure~\ref{beta_L_encoding_accuracy}B).

\begin{figure}
\centering
\includegraphics[width=.99\textwidth,trim=0mm 0mm 0mm 0mm,clip]{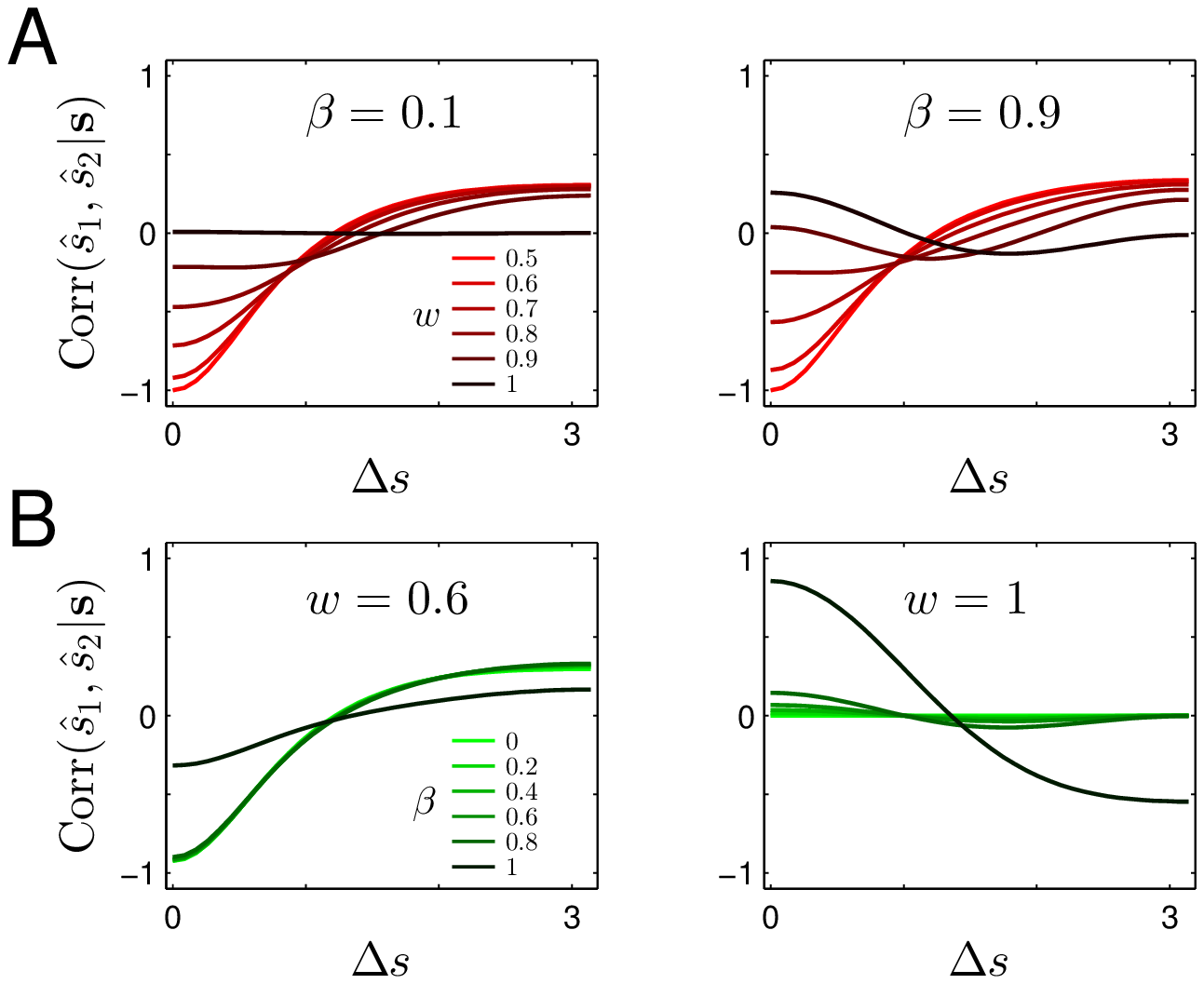} 
\caption{The asymptotic correlation between the estimates, $\mathrm{Corr}(\hat{s}_1,\hat{s}_2|\mathbf{s})$, as a function of $\Delta s$. (A) The effect of changing $w$. The left panel shows correlations for  $\beta = 0.1$ and the right panel shows correlations for $\beta = 0.9$. Different curves in each panel correspond to different $w$ values indicated in the inset in the left panel. (B) The effect of changing $\beta$. The left panel shows correlations for a strong stimulus mixing regime ($w=0.6$) and the right panel shows correlations under the no stimulus mixing regime ($w=1$). Different curves in each panel correspond to different $\beta$ values indicated in the inset of the left panel.}
\label{corr_beta_w_analytic_fig}
\end{figure}

\textbf{Correlations between the estimates:} The FIM also predicts prominent stimulus-dependent correlations between the estimates of the two stimuli. The asymptotic correlation between the optimal estimates is given by Equation~\ref{corr_analytic} ({\it Materials and Methods}). In Figure~\ref{corr_beta_w_analytic_fig}, we show the correlations between the two estimates under different parameter regimes, using the Poisson-like noise model. Figure~\ref{corr_beta_w_analytic_fig}A and \ref{corr_beta_w_analytic_fig}B show the effects of changing $w$ and $\beta$ respectively on the correlation between the estimates. Psychophysical tasks where subjects have to estimate multiple stimuli simultaneously are uncommon (see Orhan and Jacobs (2013) for an exception), but Figure~\ref{corr_beta_w_analytic_fig} suggests that the pattern of correlations between the estimates obtained from such tasks can be potentially informative about the optimality of the subjects' decoding methods and about possible parameter regimes for their encoding models.

\subsection*{A reduced model to understand the effects of stimulus mixing}
\begin{figure}
\centerline{\includegraphics[width=.99\textwidth,trim=0mm 0mm 0mm 0mm,clip]{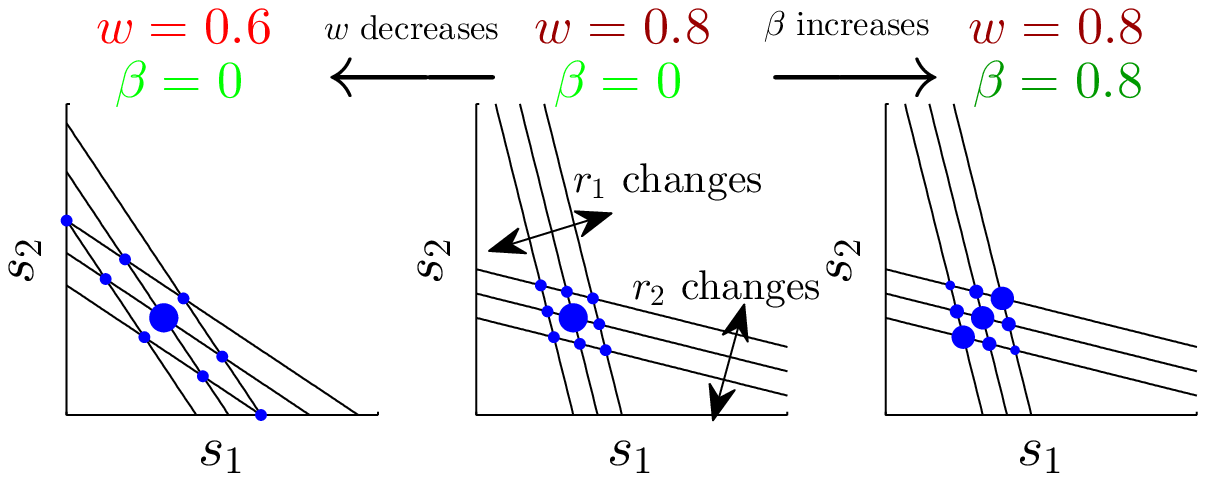}}
\caption{Geometric intuition for the effects of $w$ and $\beta$ on the variance of estimates in a simplified two-neuron model (Equation~\ref{reduced_model_linear_eq}). Maximum-likelihood estimates of the stimuli are represented by the blue dots, as $r_1$ and $r_2$ vary stochastically for a particular stimulus pair $(s_1,s_2)$. Dot size represents the probability of the corresponding estimates.}\label{reduced_model_linear}
\end{figure}

To develop a geometric intuition for the effects of stimulus mixing and across-group neural correlations on encoding accuracy, we consider a simpler, reduced version of the linear mixing model. In this model, neurons in each group are reduced to a single neuron. In addition, we model the responses of these two ``reduced neurons'' linearly, ignoring the non-linearity introduced by the tuning function, $f$. Thus, the responses of the two neurons are modeled as:
\begin{equation}
r_1 = w s_1 + (1-w)s_2 + \varepsilon_1, \qquad r_2 = (1-w) s_1 + w s_2 + \varepsilon_2 , \label{reduced_model_linear_eq}
\end{equation}
where $\varepsilon_1$ and $\varepsilon_2$ are zero-mean random variables with correlation $\beta$, representing correlated noise in the responses. A given $(r_1,r_2)$ pair describe two lines in the $(s_1,s_2)$ plane: $r_1 = w s_1 + (1-w)s_2$ and $r_2 = (1-w) s_1 + w s_2$. The maximum likelihood estimate of the stimuli is given by the intersection of these two lines. As $r_1$ and $r_2$ vary stochastically from trial to trial due to noise, the lines as well as their intersection point change. If there is any stimulus mixing ($w\neq 1$), the geometry of the lines dictates that the estimates should be stretched along the anti-diagonal direction, making them more variable than under the no mixing condition ($w=1$) for the same $(r_1,r_2)$ values. This is illustrated in the middle panel in Figure~\ref{reduced_model_linear} for $w = 0.8$ and $\beta = 0$. Increasing the stimulus mixing makes the slopes of the lines more similar to each other, which stretches the intersection points even further (left panel in Figure~\ref{reduced_model_linear}) and increases their variance. Increasing the across-group neural correlation $\beta$, on the other hand, makes the intersection points along the diagonal more probable (right panel in Figure~\ref{reduced_model_linear}), counteracting the anti-diagonal stretching caused by stimulus mixing and decreasing the variance of the estimates.

\subsection*{Stimulus dependence of the variance}
\begin{figure}
\centerline{\includegraphics[width=.99\textwidth,trim=0mm 0mm 0mm 0.0mm,clip]{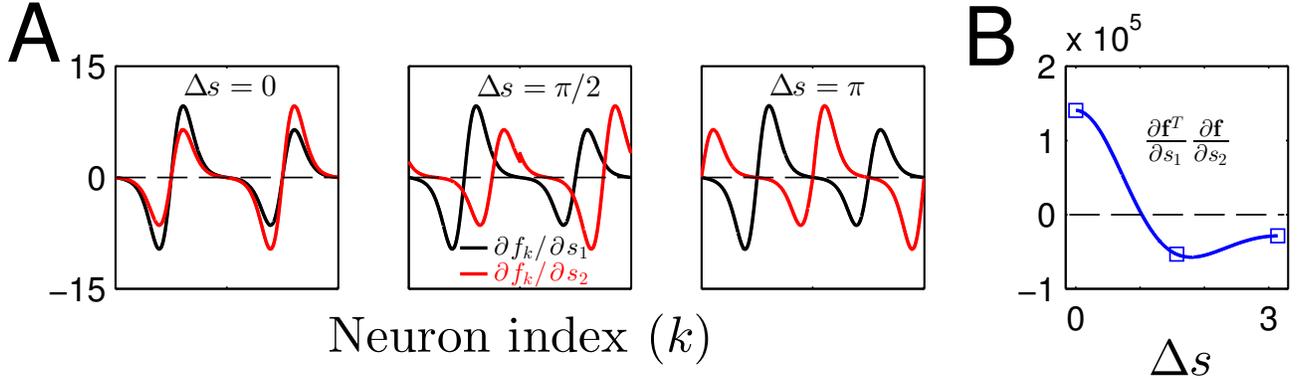}}
\caption{(A) Derivatives of the mean responses with respect to $s_1$ (black) and $s_2$ (red) plotted for three different $\Delta s$ values. For all three panels, $w=0.6$. (B) The inner product $\frac{\partial \mathbf{f}^T}{\partial s_1}\frac{\partial \mathbf{f}}{\partial s_2}$ as function of $\Delta s$ (blue line) and the points corresponding to the $\Delta s$ values shown in (A) are indicated by the open square signs.}\label{derivative_profiles}
\end{figure}

The dependence of the asymptotic variance of the optimal estimates on $\Delta s$ cannot be explained with the reduced model. {\color{black}To gain some insight into the mechanism behind this dependence, we consider the linear mixing model with additive noise and independent neurons. In this case, it can be shown that the encoding precision is proportional to (see Equation \ref{additive_ind_var} in {\it Materials and Methods}):
\begin{equation}
\Big\|\frac{\partial \mathbf{f}}{\partial s_1} \Big\|^2 - \Bigg( \frac{\frac{\partial \mathbf{f}^T}{\partial s_1} \frac{\partial \mathbf{f}}{\partial s_2}}{\|\frac{\partial \mathbf{f}}{\partial s_1} \|} \Bigg)^2 ,  \label{additive_ind_prec}
\end{equation}
where $\| \cdot \|$ denotes the Euclidean norm and $\mathbf{f}$ is a column vector of the mean responses of all neurons in both groups. In the linear mixing model $\|\frac{\partial \mathbf{f}}{\partial s_1} \|$ does not depend on $\Delta s$, hence the dependence of encoding precision on $\Delta s$ is determined solely by the magnitude of the inner product $\frac{\partial \mathbf{f}^T}{\partial s_1}\frac{\partial \mathbf{f}}{\partial s_2}$: when the magnitude of this inner product is large relative to the norms of the individual vectors, i.e. when the derivative profiles with respect to $s_1$ and $s_2$ overlap more, the encoding precision becomes low. Figure~\ref{derivative_profiles} shows that $\frac{\partial \mathbf{f}}{\partial s_1}$ and $\frac{\partial \mathbf{f}}{\partial s_2}$ overlap more extensively for smaller $\Delta s$, explaining why stimulus mixing is especially harmful for stimuli with small $\Delta s$.

Although the simple proportionality relation above does not hold in the case of Poisson-like noise (or for correlated neurons for that matter), it qualitatively captures the stimulus-dependence of the asymptotic variance in Figure~\ref{w_beta_hetero_fig}.}

\subsection*{Generalization to more than two stimuli}
It is straightforward to generalize the preceding linear mixing model to more than two stimuli. However, deriving a simplified mathematical expression for the FIM, as was done for the case of two stimuli, becomes infeasible for this case. Therefore, we present results from numerically computed FIMs for up to 6 stimuli.  

\begin{figure}
\centerline{\includegraphics[width=.99\textwidth,trim=0mm 0mm 0mm 0.5mm,clip]{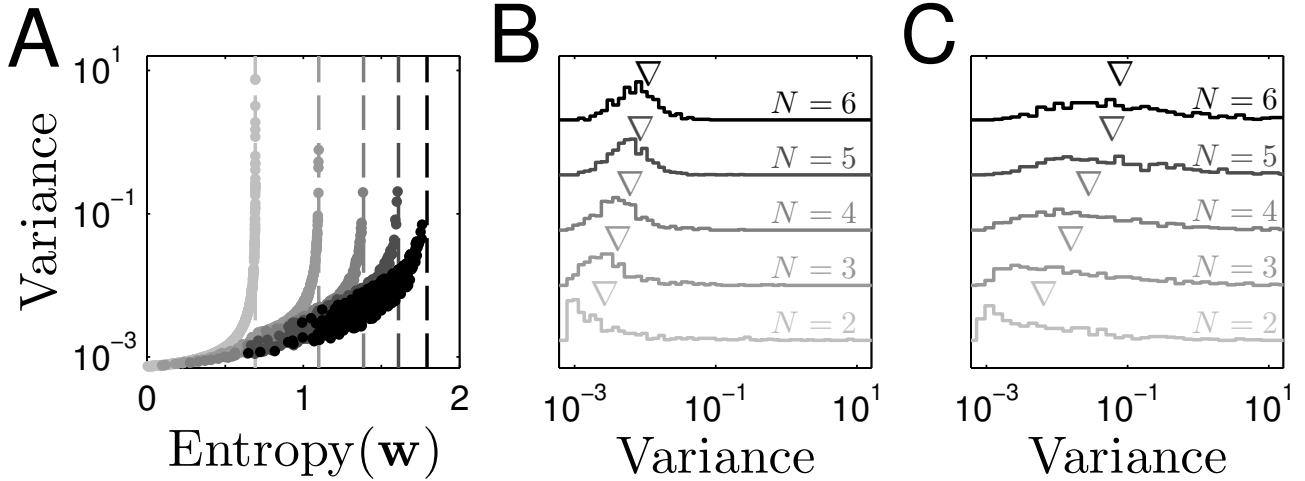}}
\caption{Linear mixing with more than two stimuli. (A) Asymptotic variance as a function of the entropy of the weight vector. Vertical dashed lines show $\log N$ (maximum entropy for each set size). (B) Histograms of the asymptotic variance of the optimal estimates for different set sizes from $N=2$ to $N=6$ with darker colors representing larger set sizes. The inverted triangles indicate the median variance for each set size. (C) Similar to (B), but the weight vectors of different groups are sampled randomly from the $(N-1)$--dimensional probability simplex. }\label{setsize_var_entropy_fig}
\end{figure}

To generate the mixing weights of different groups of neurons, for each set size $N$, we first draw a random weight vector $\mathbf{w}$, uniformly distributed on the $(N-1)$--dimensional probability simplex, i.e. the region defined by $\sum_i w_i = 1$ and $w_i\geq 0$. We then generate an $N \times N$ circulant matrix $\mathbf{W}$ from the weight vector $\mathbf{w}$. The rows of this matrix, which are all circular permutations of $\mathbf{w}$, give the weight vectors of each group in the population. We generate $512$ such weight matrices and, {\color{black}for each weight matrix}, compute the asymptotic variance of the optimal estimator from the inverse of the Fisher information matrix for the particular stimulus configuration where all the stimuli are identical, $s_1 = s_2 = \ldots = s_N$. The number of neurons per group and the magnitude of noise correlations between groups are held constant across set sizes in these simulations. Similar to the results for $N=2$, the estimation variance increases when the weight vector $\mathbf{w}$ becomes more uniform, i.e. when different groups become equally responsive to all stimuli. To quantify the uniformity of the weight vectors, we use the Shannon entropy of $\mathbf{w}$ treated as a discrete probability distribution. Figure~\ref{setsize_var_entropy_fig}A shows the asymptotic estimation variance as a function of the Shannon entropy of the weight vector. When {\color{black}the logarithm of} the estimation variance is linearly regressed on {\color{black}the logarithm of} the set size $N$, we find a highly significant effect with a positive slope of about $0.82$ ($p<.0001$), suggesting that the variance increases with set size (Figure~\ref{setsize_var_entropy_fig}B). {\color{black}This result is not sensitive to the particular way the weight matrices $\mathbf{W}$ are chosen and holds as well for the case where the weight vectors of different groups in the population, rather than being circular permutations of a single weight vector, are random samples from the $(N-1)$--dimensional probability simplex (Figure~\ref{setsize_var_entropy_fig}C). In this case, the linear regression of the logarithm of the estimation variance on the logarithm of the set size yields a highly significant effect with a positive slope of about $2.2$ ($p<.0001$), again suggesting an increase in the estimation variance with set size.} This increase is not caused by a reduction in gain per stimulus, as the number of neurons per group was held constant and the presented stimuli were identical. Rather, it is due to an increase with $N$ in the mean normalized entropy (normalized by the maximum possible entropy, i.e. $\log N$) of weight vectors drawn from a probability simplex (Nemenman et al., 2002). In other words, with increasing $N$, it becomes more and more difficult to find ``harmless'', low-entropy weight vectors. This result suggests a novel mechanism that might contribute to set size effects, i.e. declines in performance with set size, observed in various psychophysical tasks (see \textit{Discussion}).

\subsection*{Generalization to a suboptimal decoder}
The results presented so far concern the FIM which describes the asymptotic behavior of the optimal estimator. An important question is to what extent these results generalize to empirically motivated suboptimal decoders. Here, we show that the effects of stimulus mixing, heterogeneity of the mixing weights, and across-group noise correlations obtained from the analysis of the FIM generalize to a particular type of suboptimal decoder called the optimal linear estimator (OLE) (Salinas and Abbott, 1994). Because OLE is a biased estimator, we use the mean squared error (MSE) as a measure of the estimator's performance.

\begin{figure}
\centerline{\includegraphics[width=.99\textwidth,trim=0mm 0mm 0mm 0mm,clip]{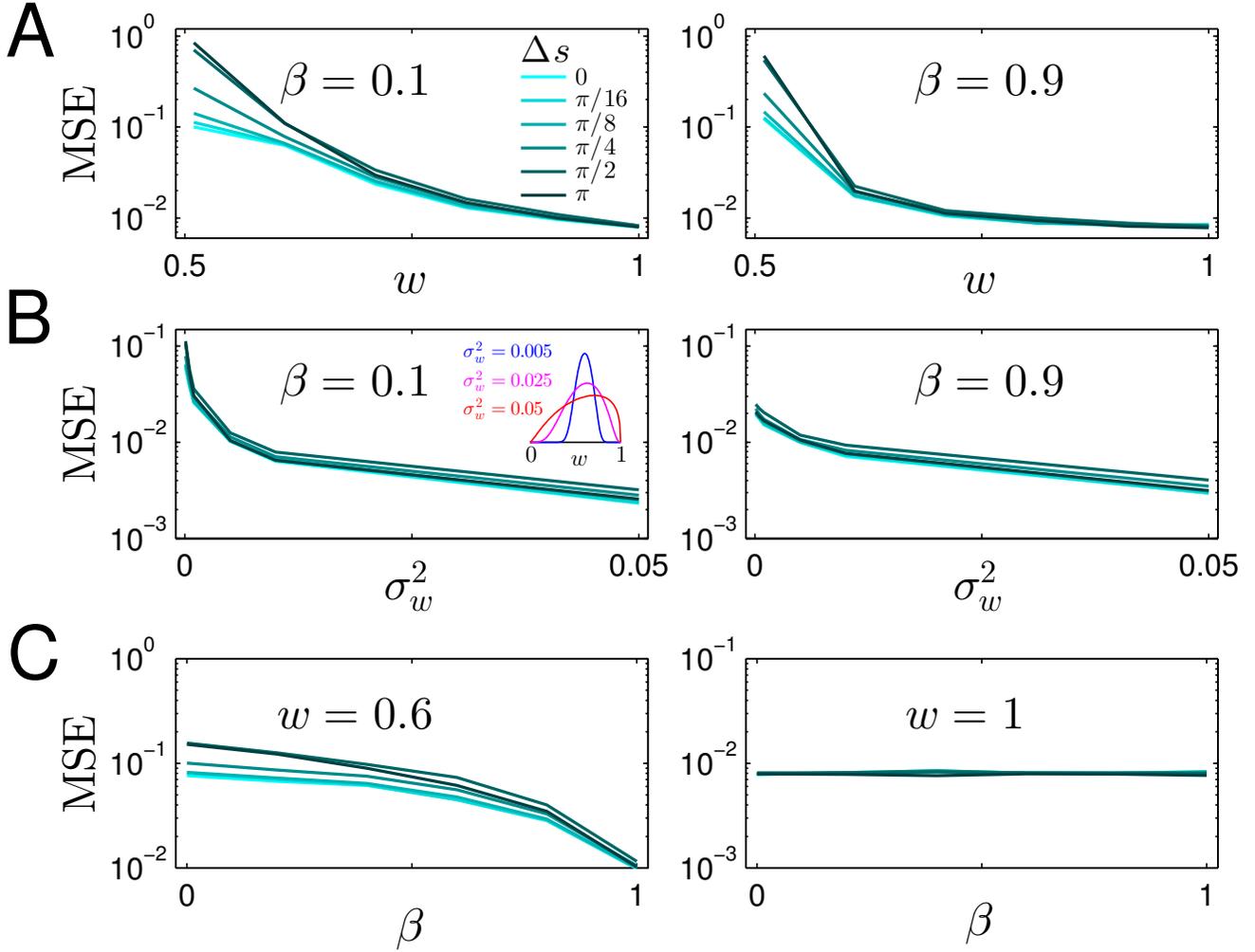}}
\caption{Results for the OLE decoder. (A) Dependence of the MSE of the optimal linear estimator on $w$. The left panel shows the MSE for a small $\beta$ value ($\beta=0.1$) and the right panel for a large $\beta$ value ($\beta=0.9$). (B) Dependence of the MSE on the heterogeneity of mixing weights. The weights $w$ are drawn from a beta distribution with variance $\sigma_w^2$. The two panels show the MSE for two different $\beta$ values. The inset in the left panel shows the weight distributions for three different $\sigma_w^2$ values. (C) Dependence of the MSE on $\beta$. The left panel shows the MSE for a small $w$ value ($w=0.6$) and the right panel shows the MSE for a large $w$ value ($w=1$). Different curves in each panel correspond to six different $\Delta s=|s_1 - s_2|$ values indicated in the inset in (A), with lighter colors corresponding to smaller $\Delta s$ values. Other parameter values are listed in {\it Materials and Methods}.} \label{ole_w_beta_hetero_fig}
\end{figure}

Figure~\ref{ole_w_beta_hetero_fig}A shows the MSE of the OLE for different degrees of stimulus mixing. Increased stimulus mixing (decreasing $w$) deteriorates the estimator's performance. This is consistent with the results presented above for the FIM. The stimulus dependence of the estimator error, however, has a different form than for the FIM. Figure~\ref{ole_w_beta_hetero_fig}B shows the MSE of the OLE for different amounts of heterogeneity in the mixing weights. Again, consistent with the results obtained from the FIM, increased heterogeneity improves the estimator's performance. Figure~\ref{ole_w_beta_hetero_fig}C shows the MSE of the OLE for different across-group correlation values. Under strong stimulus mixing ($w=0.6$), increasing $\beta$ improves the decoder's performance by up to an order of magnitude in most cases. This effect is also consistent with the results presented earlier for the FIM. The effect of $\beta$, however, becomes less significant under the no stimulus mixing condition ($w=1$). The results for the OLE are also, in general, less dependent on $\Delta s$ than the results for the FIM.

\subsection*{Nonlinear forms of stimulus mixing}
So far, we have considered a linear stimulus mixing model where the responses of neurons to multiple stimuli are modeled as simple linear combinations of their responses to individual stimuli alone. Do our results also hold for other forms of stimulus mixing, or is the assumption of linearity crucial? To show that our results are not specific to the linear mixing model, here we consider two nonlinear, experimentally motivated forms of stimulus mixing.

{\bf Nonlinear mixing model of Britten and Heuer (1999):} 
Britten and Heuer (1999) present pairs of moving gratings inside the receptive fields of MT neurons and show that a nonlinear mixing equation of the following form provides a good characterization of their responses:
\begin{equation}
f_k(s_1,s_2) = a \Big(\frac{f(s_1;\phi_k)^\nu + f(s_2;\phi_k)^\nu}{2}\Big)^{1/\nu} + b , \label{brittenheuer99_eq} 
\end{equation}
where $f(s_1;\phi_k)$ and $f(s_2;\phi_k)$ are the mean responses of neuron $k$ to the individual gratings. This equation can interpolate smoothly between simple averaging ($a=1$, $\nu=1$) and max-pooling ($\nu\rightarrow\infty$) of responses to the individual gratings. Britten and Heuer (1999) report a wide range of values for the parameters $a$ and $\nu$ across the population of neurons they recorded from. They further show that allowing these parameters to vary results in significantly better fits than the simple averaging model for most of the neurons.

\begin{figure}
\centerline{\includegraphics[width=.99\textwidth,trim=0mm 0mm 0mm 0mm,clip]{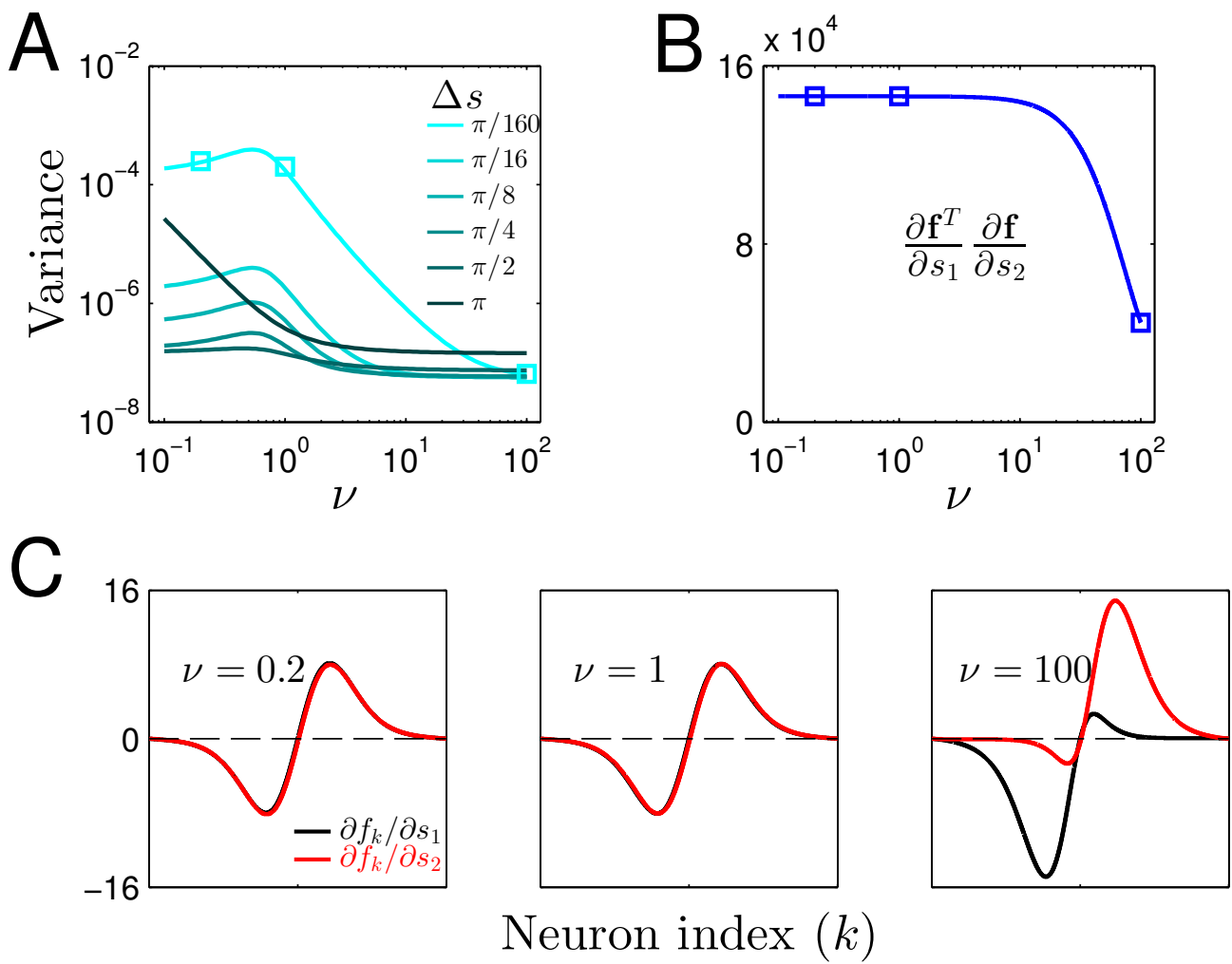}}
\caption{Analysis of the nonlinear mixing model of Britten and Heuer (1999). (A) The asymptotic variance of the optimal estimator as a function of the exponent $\nu$ (Equation~\ref{brittenheuer99_eq}). Different curves correspond to six different $\Delta s=|s_1 - s_2|$ values indicated in the inset, with lighter colors corresponding to smaller $\Delta s$ values. Note that for $\Delta s = 0$, the variance diverges in this model as in the linear mixing model with $w=0.5$. (B) The inner product $\frac{\partial \mathbf{f}^T}{\partial s_1}\frac{\partial \mathbf{f}}{\partial s_2}$ as function of $\nu$ (for $\Delta s = \pi/160$). (C) Profiles of the partial derivatives of the mean responses for three different $\nu$ values: $\nu=0.2$, $\nu=1$ and $\nu=100$. These points are also indicated by the open square signs in (A) and (B). Parameter values for the results shown here are listed in {\it Materials and Methods}.} \label{brittenheuer99_fig}
\end{figure}

We assume a single unsegregated population of neurons and derive the Fisher information matrix as before, using the mean responses described in Equation~\ref{brittenheuer99_eq} above. Figure~\ref{brittenheuer99_fig}A shows the asymptotic variance of the optimal estimator as a function of the exponent $\nu$. Increasing $\nu$ reduces stimulus mixing and significantly improves the encoding accuracy, consistent with the results from the linear mixing model. As in the linear mixing model, the effect of stimulus mixing on encoding accuracy can be understood, at least qualitatively, by considering the magnitude of the overlap between the profiles of the partial derivatives of the mean responses with respect to the two stimuli, $\frac{\partial \mathbf{f}}{\partial s_1}$ and $\frac{\partial \mathbf{f}}{\partial s_1}$. For small $\nu$ values, there is significant overlap between the partial derivative profiles, leading to a severe reduction in encoding accuracy, whereas larger $\nu$ values make the neurons sensitive to changes in only one of the stimuli and thus reduce the overlap between the derivative profiles (Figure~\ref{brittenheuer99_fig}B-C). 

{\bf Divisive stimulus mixing:} We next consider another biologically motivated form of stimulus mixing based on divisive normalization. Specifically, as in the linear mixing model, we separate the neurons into two groups and assume that the responses of neurons in each group are normalized by a weighted sum of the activity of neurons in the other group. The weighting is assumed to be neuron-specific such that neurons with similar stimulus preferences exert a larger divisive influence on each other. This type of divisive normalization has previously been motivated by considerations of efficient coding in early visual cortical areas (Schwartz and Simoncelli, 2001) and can be used to describe stimulus-dependent suppressive surround effects in the visual cortex (Allman et al., 1985; Cavanaugh et al., 2002). Mathematically, the response of a neuron in the first group is described by:    
\begin{equation}
f_k(s_1,s_2) = \frac{f(s_1; \phi_k)^2}{\Delta + k_w \sum_{k^\prime} w(\phi_k, \phi_{k^\prime}) f(s_2; \phi_{k^\prime})^2} , 
\label{div_norm_mean_resp_eq}
\end{equation}
where for the weighting profile $w(\phi_k, \phi_{k^\prime})$, we use a normalized von Mises function ({\it Materials and Methods}). Responses of neurons in the second group are similar, but with the roles of $s_1$ and $s_2$ reversed.

\begin{figure}
\centerline{\includegraphics[width=.99\textwidth,trim=0mm 0mm 0mm 0mm,clip]{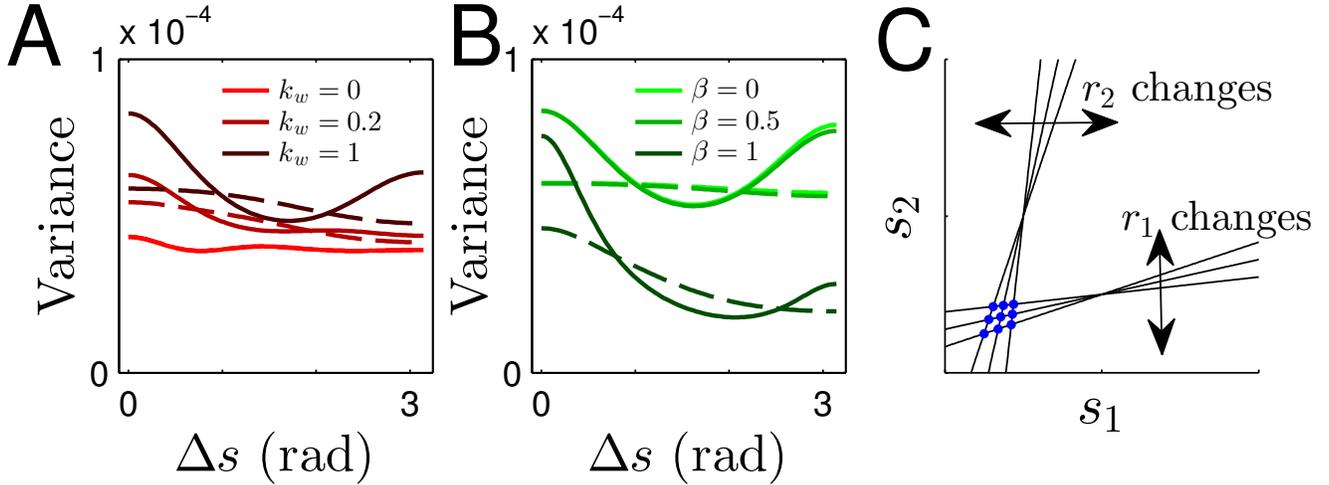}}
\caption{Analysis of the divisive mixing model. (A) The effect of varying $k_w$. Dashed lines in (A) and (B) show the results for a model with no stimulus mixing, but with gain matched neuron-by-neuron to that of the divisive mixing model. (B) The effect of varying $\beta$. Other parameter values are listed in {\it Materials and Methods}. (C) Geometric intuition for the mildly harmful effects of a divisive form of stimulus mixing in a simplified two-neuron model (Equation~\ref{toy_divisive_eqs}). Maximum-likelihood estimates of the stimuli are represented by the blue dots.} \label{divisive_all_fig}
\end{figure}

Figure~\ref{divisive_all_fig}A-B show the effects of varying the divisive normalization scaling factor $k_w$ and the across-group neural correlations $\beta$ on encoding accuracy. Increasing $k_w$ decreases encoding accuracy. However, unlike in the linear mixing model, this decrement in encoding accuracy does not only reflect the effect of stimulus mixing \textit{per se}, but also a stimulus-independent scaling of the response gain. To tease apart the contribution of stimulus mixing \textit{per se} from that of a simple gain change in the responses, we built a model that, neuron by neuron, had the same gain as the divisive normalization model in Equation~\ref{div_norm_mean_resp_eq} (for all stimuli), but whose Fisher information was computed by treating the denominator in Equation~\ref{div_norm_mean_resp_eq} as constant (this was done by setting the off-diagonal entries of the FIM to zero). This second model thus eliminates stimulus mixing, but preserves the neuron-by-neuron response gains in the first model. The results for this second model are shown with dashed lines in Figure~\ref{divisive_all_fig}A-B. Comparing the dashed lines with the same colored solid lines, we see that stimulus mixing \textit{per se} induces a cost to encoding accuracy for stimulus pairs with large and small $\Delta s$ values, but not for stimulus pairs with intermediate $\Delta s$ values. As in the linear mixing model, increasing $\beta$ improves encoding accuracy in the divisively normalized model as well (Figure~\ref{divisive_all_fig}B).

Figure~\ref{divisive_all_fig}A suggests that stimulus mixing in the divisively normalized form given in Equation~\ref{div_norm_mean_resp_eq} is not as harmful as in the linear mixing model (or in the nonlinear mixing model of Britten and Heuer). To understand this difference between the two forms of stimulus mixing, we consider a two-neuron version of the divisively normalized model analogous to the two-neuron model considered earlier for the linear mixing model. We again ignore the nonlinearities introduced by the tuning function $f(s; \phi)$ and model the responses of the two neurons as follows:
\begin{equation}
r_1 = \frac{s_1}{\Delta + w s_2} +\varepsilon_1, \qquad r_2 = \frac{s_2}{\Delta + w s_1} + \varepsilon_2 . \label{toy_divisive_eqs}
\end{equation}
As in the two-neuron version of the linear mixing model, for a given $(r_1,r_2)$ pair, the two equations above define two lines in the $(s_1,s_2)$ plane whose intersection gives the maximum likelihood estimate of the stimuli. We first note that unlike in the linear mixing model, noise in $r_i$ changes the slopes of the lines. The slopes of the two lines described in Equation~\ref{toy_divisive_eqs} are given by $1/(r_1 w)$ and $r_2 w$ respectively. This suggests that as long as $r_1$ and $r_2$ are sufficiently large, the slope of the first line will be small, and will remain small despite random variations in $r_1$, whereas the slope of the second line will be large and will remain large in the face of variations in $r_2$. Unless $r_1$ and $r_2$ are very small, the two neurons thus encode the stimuli roughly orthogonally (Figure~\ref{divisive_all_fig}C), unlike in the case of linear mixing where the slopes of the lines can become arbitrarily close to each other as stimulus mixing increases. This approximately orthogonal coding of the stimuli, in turn, causes only a relatively small amount of distortion in the estimates $(\hat{s}_1,\hat{s}_2)$ as $r_1$ and $r_2$ vary stochastically for a particular stimulus pair (as indicated by the spread of the blue dots in Figure~\ref{divisive_all_fig}C), explaining why stimulus mixing is less harmful in the divisive mixing model than in the linear mixing model. As in the linear mixing model, the stimulus dependence of encoding accuracy in the divisive mixing model can be qualitatively understood by considering the magnitude of the inner product of the derivatives of the mean responses with respect to $s_1$ and $s_2$ (not shown). 

\subsection*{Stimulus mixing is not always harmful for encoding accuracy}
The examples of stimulus mixing considered thus far showed a harmful effect on encoding accuracy. This raises the important question: is stimulus mixing always harmful for encoding accuracy? Here we show that the answer is no. To show this, we first analyze the general stimulus mixing problem with a two-dimensional toy model similar to the ones presented earlier for linear and divisive mixing. We imagine two ``neurons'' mixing the two stimuli $s_1$, $s_2$ according to $f_1(s_1,s_2)$ and $f_2(s_1,s_2)$ respectively. The responses of the two neurons are given by $r_1 = f_1(s_1,s_2) + \varepsilon_1$ and $r_2 = f_2(s_1,s_2) + \varepsilon_2$ where $\varepsilon_1$ and $\varepsilon_2$ are Gaussian random variables with variance $\sigma^2$ and correlation $\rho$. We denote the Jacobian matrix for the mean responses of the neurons by $\mathbf{J}$. One can think of $\mathbf{J}$ as a mixing matrix describing the sensitivity of each neuron to changes in each stimulus. $\mathbf{J}$ would be diagonal (or anti-diagonal) in the case of no stimulus mixing. The FIM is given by $I_F = \mathbf{J}^T \mathbf{\Sigma}^{-1}\mathbf{J}$ where $\mathbf{\Sigma}$ is the covariance matrix of the response noise. $I_F^{-1}$ gives the asymptotic covariance matrix of the maximum likelihood estimator. To find the optimal mixing matrix $\mathbf{J}$, we minimize the trace of $I_F^{-1}$, i.e. $\mathrm{Tr}[I_F^{-1}] = \mathrm{Tr}[\mathbf{J}^{-1} \mathbf{\Sigma} \mathbf{J}^{-T}]$ with respect to $\mathbf{J}$. With no constraints on $\mathbf{J}$, $I_F^{-1}$ can be made arbitrarily small, for example by making $\mathbf{J}$ diagonal and its diagonal entries arbitrarily large. Because the derivatives in the Jacobian can be negative or non-negative, a plausible constraint on $\mathbf{J}$ is to require the sum of the squares of the derivatives in $\mathbf{J}$ to be a finite constant $K$. In terms of the matrix $\mathbf{J}$, this means requiring $\mathrm{Tr}[\mathbf{J}^T \mathbf{J}] = K$. The optimal $\mathbf{J}$ can then be found by the method of Lagrange multipliers ({\it Materials and Methods}).

The optimal solution is to set the gradients of the two neurons, $\nabla f_1$ and $\nabla f_2$, to have equal norm and the angle between them to $\theta^\ast$ with $\cos \theta^\ast$ given by (Figure~\ref{optimal_mixing_fig}A):
\begin{equation}
\cos \theta^\ast = \frac{1-\sqrt{1-\rho^2}}{\rho} . \label{cos_theta_eq}
\end{equation}
The absolute orientation of the gradients in the plane, however, can be arbitrary. Thus, Equation~\ref{cos_theta_eq} describes a one-dimensional family of solutions. Because Fisher information is a local measure of information, this solution holds for a given arbitrary point $(s_1,s_2)$ in the plane. For optimal mixing over the entire plane, the conditions specified by the solution have to be satisfied for each point $(s_1,s_2)$ in the plane. This can be achieved by choosing the mixing function of the first neuron $f_1(s_1,s_2)$ arbitrarily and then choosing the mixing function of the second neuron $f_2(s_1,s_2)$ such that the optimality conditions on the gradients are satisfied at each point in the plane.

Three important aspects of the solution are worth emphasizing. First, the solution does not require $\mathbf{J}$ to be diagonal (or anti-diagonal). Thus, stimulus mixing is not intrinsically harmful, but rather stimuli should be mixed in complementary ways by different neurons so that the conditions on the gradients are satisfied. Second, the optimal solution, in fact, necessitates a certain amount of stimulus mixing when $\rho \neq 0$, as the optimality condition for $\rho \neq 0$ cannot be satisfied with $\nabla f_1$ and $\nabla f_2$ aligned with the two axes of the $(s_1,s_2)$ plane. Third, for $\rho \longrightarrow 0$, $\cos \theta^\ast \longrightarrow 0$; therefore, the gradients have to be orthogonal to each other in this case. Furthermore, the optimal angle $\theta^\ast$ between the gradients changes rather slowly as $\rho$ moves away from 0. Thus the gradients should be close to orthogonal for a large range of $\rho$ values around 0. The orthogonality condition can be understood as follows. Intuitively, $\nabla f_1$ is the first neuron's ``least uncertain'' direction in the $(s_1,s_2)$ plane. The second neuron has to align {\it its} least uncertain direction orthogonally to $\nabla f_1$ so that together the two neurons can encode the stimuli with the least total uncertainty. In our original analysis of the linear mixing model, the orthogonality condition can be satisfied only when there is no stimulus mixing, because, motivated by a consideration of consistency with physiological data, the weights were assumed to be non-negative in that case.

\begin{figure}
\centering
\includegraphics[width=.99\textwidth,trim=0mm 0mm 0mm 0mm,clip]{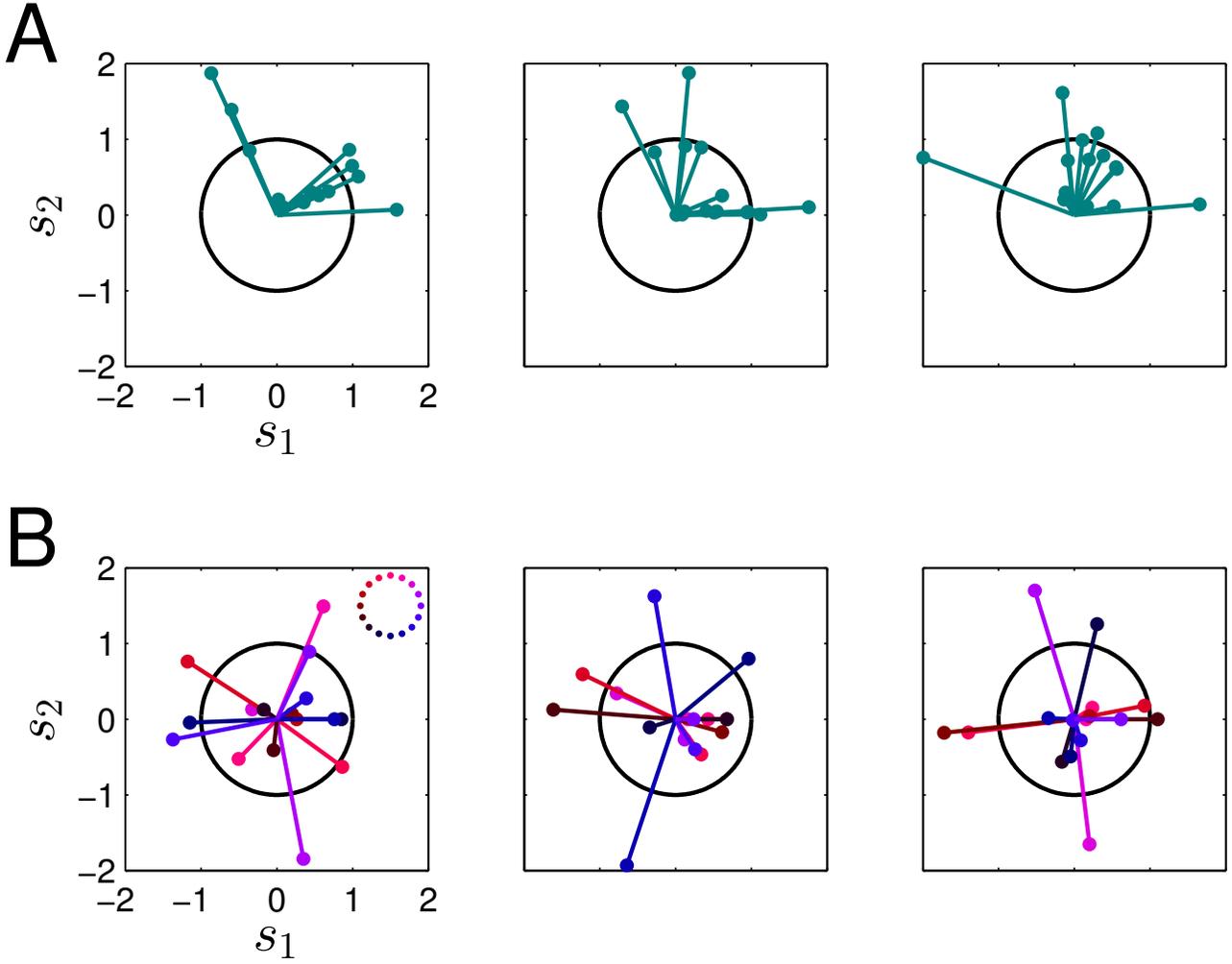} 
\caption{Numerical solution of the optimal mixing problem. (A) Three example numerical solutions to the optimal stimulus mixing problem for $n$ independent neurons. In each panel, the spokes represent the gradients of different neurons in the population and the black circle shows the unit circle. (B) Three example numerical solutions to the optimal stimulus mixing problem for $n$ correlated neurons. Correlations between the neurons are assumed to be limited-range, i.e. the correlation matrix $R$ has the form given in Equation~\ref{R_matrix} (with $c_0=0.3$ and $L=2$). The noise is assumed to be additive Gaussian with constant standard deviation $\sigma=1$ for all neurons. In (B), the gradient vectors are colored such that neurons with higher correlations have more similar colors, where the similarity of the colors is indicated by their position on the circle shown in the inset of the left panel. The population has $n=16$ neurons in the examples shown here.}
\label{three_solutions_fig}
\end{figure}

\begin{figure}
\centerline{\includegraphics[width=0.99\textwidth,trim=0mm 0mm 0mm 0mm,clip]{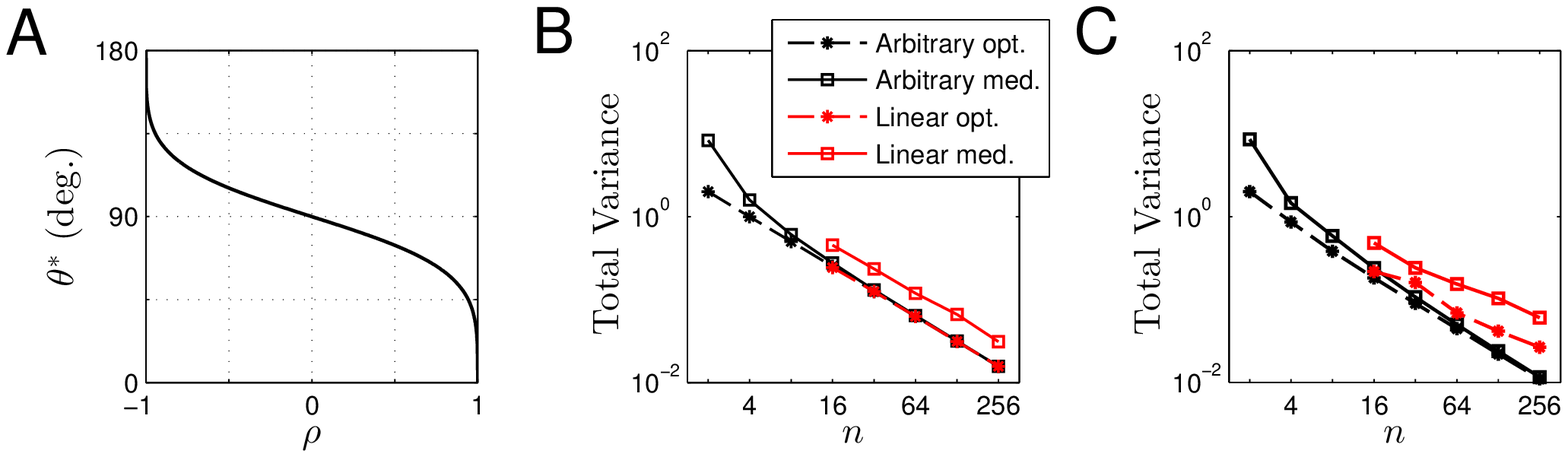}}
\caption{Optimal stimulus mixing. (A) Optimal angle between the gradients as a function of the correlation $\rho$ in a two-dimensional stimulus mixing problem. (B) Total variance, i.e. $\mathrm{Tr}[I_F^{-1}]$, for the median random model compared with the total  variance for the optimal model for both linear and arbitrary encoding models. For the linear mixing model, results for $n \geq 16$ are shown only. With smaller $n$, the results are highly sensitive to the choice of tuning function centers. For random models, medians are computed over 256 random realizations of the encoding model. (C) Similar to (B), but for correlated neurons ($c_0 = 0.3$ for both the arbitrary encoding model and the linear encoding model and $\beta=0.9$ for the linear encoding model).}\label{optimal_mixing_fig}
\end{figure}

The solution of the two-dimensional toy model can be readily generalized to models with more than two neurons under certain conditions (Equations~\ref{linear_mixing_eq_first}-\ref{resource_K}; see {\it Materials and Methods}). For the general case of $n$ neurons encoding two stimuli, as far as we know, there is no closed-form solution for the optimal mixing matrix, $\mathbf{J}$, subject to a constraint on the total power of the derivatives. Numerical solution of this more general problem shows that for any given neural covariance structure, there is a diverse set of solutions: Figure~\ref{three_solutions_fig}A shows three example solutions for $n=16$ independent neurons and Figure~\ref{three_solutions_fig}B shows three example solutions for $n=16$ correlated neurons with a limited-range correlation structure. Moreover, random mixing of the stimuli by neurons performs remarkably well especially for large $n$. Figure~\ref{optimal_mixing_fig}B compares the performance of the median random mixing model with that of the optimal mixing model for different $n$ (compare the black asterisks vs. the black open squares). For the random mixing models, the gradients of neurons were chosen subject to the total power constraint, i.e. the sum of the squared norms of gradients was constant, but they were otherwise random.

We also wanted to see how well linear mixing models perform compared to unconstrained encoding models where the gradients can be set arbitrarily (subject to the resource constraint). When the encoding model is constrained to be a linear mixing model with two groups and fixed weights for each stimulus (Equations~\ref{linear_mixing_eq_first}-\ref{linear_mixing_eq}; see {\it Materials and Methods}), random ensembles of linear mixing models perform worse than random ensembles of arbitrary encoding models (compare the red vs. black open squares in Figure~\ref{optimal_mixing_fig}B). Interestingly, however, the optimal solution for the linear mixing model appears to have the same form as the optimal solution of the two-dimensional problem and performs as well as the optimal solution for arbitrary encoding models (compare the red vs. black asterisks in Figure~\ref{optimal_mixing_fig}B). Figure~\ref{optimal_mixing_fig}B shows the results for populations with independent neurons. Analogous results hold for correlated neural populations. With correlated neurons, the improvement in the relative performance of random ensembles with increasing $n$ becomes somewhat slower and the optimal linear encoding model no longer achieves the same performance as the optimal arbitrary encoding model (Figure~\ref{optimal_mixing_fig}C).

\section*{Discussion}
Theoretical studies of population coding have traditionally focused on the encoding of single stimuli by neural populations (Abbott and Dayan, 1999; Sompolinsky et al., 2001; Ecker et al., 2011). In this work, we extended the neural population coding framework to the encoding of multiple stimuli, assuming encoding models with biologically motivated properties. We examined a linear mixing rule commonly observed in cortical responses. According to this rule, the response to the presentation of multiple objects can be described as a linear combination of the responses to the constituent objects. We find that this rule incurs a severe cost to encoding accuracy. This cost is directly related to the mixing of the stimuli in the neural responses, is independent of any general decrease in the overall activity of the population, and can be larger than the cost incurred by even large reductions in the gain or large increases in neural variability. As noted earlier, this result could explain why attention acts primarily as a stimulus selection mechanism (Reynolds et al., 1999; Pestilli et al., 2011), rather than purely as a gain increase or noise reduction mechanism. However, it should be emphasized that mechanistically, stimulus selection can be implemented with a combination of gain increase and a nonlinearity that would amplify the gain differences (Reynolds and Heeger, 2009; Pestilli et al., 2011). Therefore, gain increases might be an integral part of the stimulus selection mechanism by which attention operates.   

Why does linear mixing seem to be so prevalent in cortical responses, if it indeed incurs such a large cost? It is possible that linear, or close-to-linear, mixing is an inevitable consequence of computations carried out for other purposes, such as achieving object representations invariant to identity-preserving transformations (Zoccolan et al., 2005; Zoccolan et al., 2007). We emphasize that our framework evaluates an encoding model in terms of the ability of a downstream decoder to accurately estimate the constituent stimuli in the face of neural variability. If, however, the goal of the computation is not accurate representation of the constituent stimuli themselves, but computing a possibly complex function of them, then linear mixing or similar models are not necessarily harmful. For example, in the problem we considered in this paper, if the computational goal were only to estimate a weighted average of the two stimuli $s_1$ and $s_2$, linearly mixed responses would be ideally suited for such a task. Finding the optimal neural codes for the representation of multiple stimuli that achieve the simultaneous objectives of successful performance in behaviorally relevant tasks (e.g., see Salinas, 2006) and accurate encoding of constituent stimuli could be an important future direction.

The harmful effects of stimulus mixing can be partially alleviated by increased across-group neural correlations or by increased heterogeneity in the mixing weights of the neurons. Importantly, all our main results concerning the linear mixing model, i.e. the effects of stimulus mixing, across-group neural correlations and heterogeneity in mixing weights generalize to the suboptimal OLE decoder. This is not a trivial result, because there is, in general, no guarantee that manipulating the properties of a neural population should affect the performance of optimal and suboptimal decoders in similar ways. Indeed, a previous study (Ecker et al., 2011), for instance, found that in the presence of diversity in neural tuning properties, limited-range correlations can be beneficial for accurately encoding a single stimulus, but this holds only if the responses are decoded optimally, and does not generalize to the suboptimal OLE decoder. 

In the linear mixing model, increasing the number of stimuli makes stimulus mixing even costlier. This result suggests that stimulus mixing might contribute to set size effects commonly observed in visual short-term memory and other psychophysical tasks. Decreases in performance with set size in such tasks are typically attributed to a capacity limitation, e.g. an upper limit on the total amount of neural activity, which might be implemented by divisive normalization (Ma et al., 2014). However, our results demonstrate that even without any constraint on the total amount of neural activity (indeed, in our simulations, total activity was proportional to set size), set size effects would be expected in the linear mixing regime, as it becomes more difficult to find harmless, low (normalized) entropy weight vectors with increasing set size. Such weight vectors can still be found through learning, but any learning algorithm would take longer to reach these low entropy regions in the weight space and it would require more fine-tuning to keep the weights in a low entropy region once such a region is found through learning, as any noise in the weights, or in the learning algorithm itself, would be more likely to push the weights out of the low-entropy region.

The property that makes linear mixing particularly harmful for encoding accuracy is not the linearity of response mixing {\it per se}. It is rather the degree of overlap, or similarity, between the derivative profiles of the neural responses with respect to different stimuli that, to a first approximation, determines how harmful a particular form of stimulus mixing can be (e.g. see Figure~\ref{derivative_profiles} and Figure~\ref{brittenheuer99_fig}B-C). Indeed, our results for the non-linear mixing rule of Britten and Heuer (1999) show that stimulus mixing can lead to a severe reduction in encoding accuracy even when mixing takes a strongly non-linear form.

Stimulus mixing, in itself, is not always harmful for encoding accuracy. As our analytic solution to the optimal mixing problem in a toy model and numerical solutions in more complex cases suggest, it may even be optimal in the presence of neural correlations. Stimulus mixing has to satisfy certain conditions in order to be unharmful for encoding accuracy. In a simple two-dimensional problem and with sufficiently low neural correlations, those conditions can be condensed into an intuitive orthogonality constraint on the gradients of the two group's mean responses. In the linear mixing model, this constraint is satisfied only if either there is no stimulus mixing at all, or negative weights are allowed. We also found that random mixing by individual neurons, assuming that there is no restriction to non-negative weights, performs remarkably well, especially in large populations. This result is reminiscent of other cases where random solutions have been found to perform well (Rigotti et al., 2010; Barak et al., 2013) and calls for a more general account of the effectiveness of such random solutions in diverse computational problems in neuroscience.

A stimulus mixing problem similar to the one investigated in this paper has been studied previously for temporal signals (White et al., 2004; Ganguli et al., 2008). Stimulus mixing in this context refers to the mixing of signals at different time points in the responses of a recurrently connected dynamical population of neurons. White et al. (2004) show that neural networks with an orthogonal connectivity matrix can achieve optimal estimation performance for temporal signals uncorrelated across time. We note that this is formally similar to the solution for the optimal mixing matrix in our toy stimulus mixing problem, where we found that orthogonal mixing matrices are optimal when neurons are independent.

Finally, our results suggest that psychophysical tasks that require the simultaneous encoding of multiple stimuli can be informative about the brain's decoding strategies. Although behavioral tasks with a single encoded stimulus can already provide useful information about the brain's decoding schemes (Hohl et al., 2013; Haefner et al., 2013), tasks with multiple encoded stimuli can yield additional and possibly richer information about the decoder. For example, an optimal decoder predicts specific types of correlations between the estimates of two stimuli in a task where both stimuli have to be estimated simultaneously (Figure~\ref{corr_beta_w_analytic_fig}). If the pattern of correlations observed in such an estimation task is found to be inconsistent with the predicted correlations from an optimal decoder, this may be taken as evidence against optimal decoding. Similarly, different types of decoders make different predictions about the stimulus dependence of encoding accuracy even in tasks that require the estimation of a single target stimulus among multiple stimuli (e.g. compare Figure~\ref{w_beta_hetero_fig} vs. Figure~\ref{ole_w_beta_hetero_fig} for the stimulus dependence of encoding accuracy using the optimal and OLE decoders, respectively). Again, such differences can be used to rule in or rule out certain decoding schemes as plausible decoding strategies the brain might be using in a given psychophysical task.


\begin{thebibliography}{99}
\bibliographystyle{apa}

\bibitem{abbottdayan1999}
Abbott LF, Dayan P (1999) The effect of correlated variability on the accuracy of a population code. Neural Comput 11:91-102.

\bibitem{allmanetal1985}
Allman J, Miezin F, McGuinness E (1985) Stimulus specific responses from beyond the classical receptive field: neurophysiological mechanisms for local-global comparisons in visual neurons. Annu Rev Neurosci 8:407-30. 

\bibitem{baraketal2013}
Barak O, Rigotti M, Fusi S (2013) The sparseness of mixed selectivity neurons controls the generalization-discrimination trade-off. J Neurosci 33:3844-56. 

\bibitem{beckkastner2007}
Beck DM, Kastner S (2007) Stimulus similarity modulates competitive interactions in human visual cortex. J Vis 7:1-12. 

\bibitem{brittenheuer1999}
Britten KH, Heuer HW (1999) Spatial summation in the receptive fields of MT neurons. J Neurosci 19:5074-84.

\bibitem{busseetal2009}
Busse L, Wade AR, Carandini M (2009) Representation of concurrent stimuli by population activity in visual cortex. Neuron 64:931-42.

\bibitem{cavanaughetal2002}
Cavanaugh JR, Bair W, Movshon JA (2002) Selectivity and spatial distribution of signals from the receptive field surround in macaque V1 neurons. J Neurophysiol 88:2547-2556. 

\bibitem{cohenkohn2011}
Cohen MR, Kohn A (2011) Measuring and interpreting neuronal correlations. Nat Neurosci 14:811-9. 

\bibitem{duncan2001}
Duncan J (2001) An adaptive coding model of neural function in prefrontal cortex. Nat Rev Neurosci 2:820-9. 

\bibitem{eckeretal2011}
Ecker AS, Berens P, Tolias AS, Bethge M (2011) The effect of noise correlations in populations of diversely tuned neurons. J Neurosci 31:14272-83. 

\bibitem{gangulietal2008}
Ganguli S, Huh D, Sompolinsky H (2008) Memory traces in dynamical systems. PNAS 105:18970-5.

\bibitem{haefneretal2013}
Haefner RM, Gerwinn S, Macke JH, Bethge M (2013) Inferring decoding strategies from choice probabilities in the presence of correlated variability. Nat Neurosci 16:235-42.

\bibitem{hohletal2013}
Hohl SS, Chaisanguanthum KS, Lisberger SG (2013) Sensory population decoding for visually guided movements. Neuron 79:167-79.

\bibitem{maetal2014}
Ma WJ, Husain M, Bays PM (2014) Changing concepts of working memory. Nat Neurosci 17:347-56.

\bibitem{macevoyepstein2009}
MacEvoy SP, Epstein RA (2009) Decoding the representation of multiple simultaneous objects in human occipitotemporal cortex. Curr Biol 19:943-947. 

\bibitem{macevoyetal2009}
MacEvoy SP, Tucker TR, Fitzpatrick D (2009) A precise form of divisive suppression supports population coding in primary visual cortex. Nat Neurosci 12:637-45. 

\bibitem{nandyetal2013}
Nandy AS, Sharpee TO, Reynolds JH, Mitchell JF (2013) The fine structure of shape tuning in area V4. Neuron 78:1102-15.

\bibitem{nemenmanetal2002}
Nemenman I, Shafee F, Bialek W (2002) Entropy and inference, revisited. In T. G. Dietterich, S. Becker, and Z. Ghahramani, editors, Adv Neural Inf Proc Syst 14, Cambridge, MA, MIT Press.

\bibitem{orhanjacobs2013}
Orhan AE, Jacobs RA (2013) A probabilistic clustering theory of the organization of visual short-term memory. Psychol Rev 120:297-328.

\bibitem{pestillietal2011}
Pestilli F, Carrasco M, Heeger DJ, Gardner JL (2011) Attentional enhancement via selection and pooling of early sensory responses in human visual cortex. Neuron 72:832-846. 

\bibitem{recanzoneetal1997}
Recanzone GH, Wurtz RH, Schwarz U (1997) Responses of MT ans MST neurons to one and two moving objects in the receptive field. J Neurophysiol 78:2904-2.

\bibitem{reynoldsetal1999}
Reynolds JH, Chelazzi L, Desimone R (1999) Competitive mechanisms subserve attention in macaque areas V2 and V4. J Neurosci 19:1736-53. 

\bibitem{reynoldsheeger2009}
Reynolds JH, Heeger DJ (2009) The normalization model of attention. Neuron 61:168-85. 

\bibitem{rigottietal2010}
Rigotti M, Ben Dayan Rubin D, Wang XJ, Fusi S (2010) Internal representation of task rules by recurrent dynamics: the importance of the diversity of neural responses. Frontiers in Computational Neuroscience 4:24. 

\bibitem{rigottietal2013}
Rigotti M, Barak O, Warden MR, Wang XJ, Daw ND, Miller EK, Fusi S (2013) The importance of mixed selectivity in complex cognitive tasks. Nature 497:585-90. 

\bibitem{salinas2006}
Salinas E (2006) How behavioral constraints may determine optimal sensory representations. PLoS Biology: 4(12):e387. 

\bibitem{salinasabbott1994}
Salinas E, Abbott LF (1994) Vector reconstruction from firing rates. J Comput Neurosci 1:89-107. 

\bibitem{schwartzsimoncelli2001}
Schwartz O, Simoncelli E (2001) Natural signal statistics and sensory gain control. Nat Neurosci 4:819-825. 

\bibitem{sompolinskyetal2001}
Sompolinsky H, Yoon H, Kang K, Shamir M (2001) Population coding in neuronal systems with correlated noise. Phys Rev E 64:051904.

\bibitem{wardenmiller2007}
Warden MR, Miller EK (2007) The representation of multiple objects in prefrontal neuronal delay activity. Cereb Cortex 17:41-50. 

\bibitem{wardenmiller2010}
Warden MR, Miller EK (2010) Task-dependent changes in short-term memory in the prefrontal cortex. J Neurosci 30:15801-10. 

\bibitem{whiteetal2004} 
White OL, Lee DD, Sompolinsky H (2004) Short-term memory in orthogonal neural networks. Phys Rev Lett 92:148102.

\bibitem{zoccolanetal2005} 
Zoccolan D, Cox DD, DiCarlo, JJ (2005) Multiple object response normalization in monkey inferotemporal cortex. J Neurosci 25:8150-8164.

\bibitem{zoccolanetal2007}
Zoccolan D, Kouh M, Poggio T, DiCarlo JJ (2007) Trade-off between object selectivity and tolerance in monkey inferotemporal cortex. J Neurosci 27:12292-12307.

\end{thebibliography}
\end{document}